# SOWAHA as a Cancer Suppressor Gene Influence Metabolic Reprogramming

Xiaohong Yi[a], Xianwen Zhang, Claire H. Zhao[b], Yuhui Chen[a], Lijun Huang[a], Hua Zhong[a]*, Yumei Wang[a]*

[a] School of Basic Medical Sciences, Chengdu University of Traditional Chinese Medicine, Chengdu, 611137, China

[b] Del Norte High School, San Diego, California, 92127, USA

*Corresponding authors: Hua Zhong(zhonghua@cdutcm.edu.cn), Yumei Wang (yumeiwang@ cdutcm.edu.cn)

## Abstract

**Background:** *SOWAHA*, also known as *ANKRD43*, is related to embryonic development and is enriched in brain tissue, neuron, and intestinal enterocytes. Studies have indicated that *SOWAHA* is associated with the microsatellite instability (MSI) status in colorectal cancer patients and could serve as a prognostic biomarker in colorectal cancer and pancreatic cancer. However, there are rarely reports about *SOWAHA* in other types of cancer and the specific mechanism of *SOWAHA* in cancer is also not well understood.

**Methods:** Based on National Center for Biotechnology Information (NCBI), The Cancer Genome Atlas (TCGA), Genotype-Tissue Expression Project (GTEx), cBioPortal, Human Protein Atlas (HPA), Integrative Onco Genomics (IntOGen), Simple Modular Architecture Research Tool (SMART), TIMER2, MethSurv, and TISIDB database, we uncovered the potential tumor genomic features of *SOWAHA*, including the correlation with prognosis, gene mutation, immune cell infiltration, and DNA methylation, and evaluated the association with tumor heterogeneity and stemness, chemokines chemokine receptors, and immunomodulators in pan-cancer. Besides, we knocked down *SOWAHA* in SW620 cells and performed RNA-seq analysis, then we conducted functional enrichment to uncover the biological significance of *SOWAHA*.

**Results:** *SOWAHA* demonstrates early diagnostic potential, with low expression levels linked to poor prognosis in GBMLGG, KIPAN, READ, and other cancers. It is associated with immune cell infiltration, DNA methylation, tumor heterogeneity, and stemness in several epithelial cancers. Furthermore, our experiments revealed that *SOWAHA* is significantly enriched in the metabolic related pathways in cancer, such as 'P53', 'PI3K-AKT signaling pathway', and 'pathways in cancer', 'Ubiquitin-Mediated Proteolysis', 'RNA Degradation', 'Sulfur Metabolism', 'Nucleotide Excision Repair'. We also identified two potential transcription factors regulating *SOWAHA*, TBX4 and FOXP2, which are misregulated in SW620 cells.

**Conclusion:** *SOWAHA*, identified as a suppressor gene, may also serve as a valuable biomarker for diagnosis, prognosis, and immunotherapy in various carcinomas, particularly colorectal carcinoma. Its role in the progression of colorectal cancer is primarily mediated through metabolic reprogramming mechanisms.

## 1. Introduction

Cancer has become one of the major global health issues. In 2022, there were nearly 20 million new cancer cases and 9.7 million cancer-related deaths[1]. The ongoing rise in incidence and mortality rates of cancer poses a significant burden on public health. In recent years, targeted therapy and immunotherapy have emerged as new fields in cancer treatment. Targeted therapy aims to inhibit tumor growth and spread by targeting specific molecular markers in cancer cells, while minimizing damage to normal cells[2]. Immunotherapy, on the other hand, works by activating the patient's own immune system to recognize and attack cancer cells, with new techniques such as immune checkpoint inhibitors and CAR-T cell therapy showing significant clinical efficacy[3]. Although targeted therapy and immunotherapy have provided new hope in cancer treatment, they still face challenges such as drug resistance, side effects, heterogeneity, and high costs[4, 5]. To address these issues, continued research is needed to explore new targets and immunotherapy strategies, while also improving existing treatment methods to enhance patient outcomes and quality of life.

*SOWAHA* is a protein-coding gene, with the full name Sosondowah Ankyrin Repeat Domain Family Member A, also known as *ANKRD43*. *SOWAHA* belongs to the ankyrin repeat domain family, which is known for its ability to regulate protein-protein interactions, playing a key role in cell signaling, cell cycle control, cell adhesion, and other important biological processes[6].

In recent years, studies have indicated that *SOWAHA* is associated with the MSI status in colorectal cancer patients and could serve as a prognostic biomarker for predicting the efficacy of immunotherapy in colorectal carcinoma[7]. There are also reports showing that *SOWAHA* is associated with prognosis in pancreatic carcinoma and oral squamous cell carcinoma[8, 9]. However, there are no reports on *SOWAHA* in other cancers. The specific mechanism of action of *SOWAHA* in cancer is also not well understood.

We conducted an extensive pan-cancer analysis of *SOWAHA*, investigating its aberrant expression, DNA mutations, DNA methylation, immune infiltration including chemokines and receptors, and its association with immunomodulators. We elucidated the significant influence of *SOWAHA* on tumor prognosis and its capacity to impact the immune response. Remarkably, we also knocked down *SOWAHA* in SW620 cell line and then performed RNA-seq, differentially expressed genes (DEGs) were greatly enriched in many enzymes activity and metabolism processes, and negatively related with pathways of metabolic reprogramming in cancer, such as 'P53 signaling pathways', 'PI3K-AKT signaling pathway,' 'pathways in cancer', 'Ubiquitin-Mediated Proteolysis', 'RNA Degradation', 'Sulfur Metabolism', 'Nucleotide Excision Repair'. Besides, CCK-8 assay results showed that the cells proliferation and viability in si*SOWAHA* groups is better than siNC groups. These discoveries underscore the potential of *SOWAHA* as a cancer suppressor gene, whose dysregulation may influence the development of carcinoma.

## 2. Materials and Methods
### 2.1. Data collection
Gene expression data, clinical data, mutation data, and methylation data of *SOWAHA* for 45 types of cancer and their corresponding normal tissues were downloaded in UCSC Xena database (https://xena.ucsc.edu/)[10], TCGA (https://www.cancer.gov/tcga)[11], GTEx database (https://www.genome.gov/Funded-Programs/genotype-Tissue-Expression-Project)[12], cBioPortal database (http://www.cbioportal.org/)[13], and Human Protein Atlas

(https://www.proteinatlas.org/)[14, 15]. The full names and abbreviations of the 45 cancer types are shown in Table 1.

**2.2 Differential Expression Analysis of *SOWAHA***
The mRNA expression of *SOWAHA* in various cancer tissues, different clinical stages, and paired adjacent normal tissues was investigated using the Sangerbox platform (http://sangerbox.com/home.html)[16], excluding cancers with fewer than three samples per cancer type. Differential expression results were obtained using Wilcoxon tests. Additionally, the open-access proteomic tool, The Human Protein Atlas, was employed to analyze the protein expression of *SOWAHA* across different tumor types, with representative immunohistochemical images of *SOWAHA* provided using the antibody (catalog number: HPA064621; Atlas Antibodies, Sigma–Aldrich).

**2.3 Prognosis Analysis of *SOWAHA***
First, gene expression and clinical data were obtained from TCGA and GTEx. Then univariate survival analysis was performed using the Cox proportional hazards model to assess the relationship between gene expression and Overall survival (OS), Disease-specific survival (DSS), and Progression-free interval (PFI) using 'survival' R package. Hazard ratios (HR) and 95% confidence intervals (CIs) was calculated for each outcome. Kaplan-Meier survival curves were generated and performed log-rank tests to evaluate significance between high expression groups and low gene expression groups. Visualization of results were generated using 'survfit' and 'ggsurvplot' R package.

**2.3 Genetic alterations analysis of *SOWAHA***
The IntOGen (IntOGen - Cancer Mutations Browser) tool[17] was utilized to obtain the distribution of the observed mutations along the mutation related amino acid sequence. The cBioPortal (cBioPortal for Cancer Genomics) was employed to investigate the mutation of *SOWAHA* across different cancer types. The Copy number alteration (CNA) of *SOWAHA* in pan-cancer was observed and analysis by Sangerbox.

**2.4 Correlation Analysis Between *SOWAHA* and High-Frequency Mutation Genes**
The correlation between *SOWAHA* expression and high-frequency mutation genes expression was assessed using the Cancer Exploration module of TIMER2.0 (http://timer.cistrome.org/)[18]. Specifically, we evaluated the relationship between *SOWAHA* and the expression of wild-type (WT) and mutated forms of EGFR, PIK3CA, MLH1, MSH2, PMS2, MSH4, MSH6, TP53, and KRAS across various cancers.

**2.5 Analysis of *SOWAHA* DNA Methylation**
The SMART platform (http://www.bioinfo-zs.com/smartapp/) [19]was used to analyze the correlation between *SOWAHA* expression and DNA methylation. We utilized the 'CpG-aggregated methylation' module to calculate the *SOWAHA* methylation levels in pan-cancer and a box plot represents the results. The MethSurv web tool (https://biit.cs.ut.ee/methsurv/) [20]was used to perform multivariable survival analysis based on DNA methylation data.

**2.6 Correlation analysis of *SOWAHA* and the tumor immune microenvironment (TIME)**
The correlation of *SOWAHA* expression and TIME can be analyzed using the ESTIMATE

(Estimation of STromal and Immune cells in MAlignant Tumor tissues using Expression data) algorithm. The ESTIMATE algorithm is designed to predict the presence of stromal and immune cells in tumor tissues using gene expression data, and it generates three key scores: stroma, immune, and estimate scores. The 'estimate" R package and Spearman correlation test were used to calculate stroma, immune, and estimate scores for each patient in each tumor based on gene expression. The results were then visualized using the 'ggplot2' R packages.

**2.7 Correlation analysis of *SOWAHA* with immune infiltration cells**

We used the TISIDB web tool (http://cis.hku.hk/TISIDB/search.php)[21] to analyze the effect of *SOWAHA* expression on tumor-infiltrating immune cells (TIICs) across various cancers. The correlation between *SOWAHA* expression and infiltration degree of 28 immune cell types was explored using Spearman correlation analysis, including active T cell(Act$^{CD8+}$, Act$^{CD4+}$), Central Memory T cell (Tcm$^{CD8+}$, Tcm$^{CD4+}$), Effector Memory T cell (Tem$^{CD8+}$, Tem$^{CD4+}$), regulatory T cell (Treg), Th1, Th2, active B cell(Act B), immature B cell (Imm B), memory B cell(Mem B), natural killer(NK), CD56(bright/dim) NK, Myeloid dendritic cell, natural killer (NK), natural killer T cell(NKT), Myeloid-derived suppressor cell (MDSC), active Dendritic Cell (Act DC), plasmacytoid dendritic cells (pDC), immature Dendritic Cell (iDC), macrophage, monocyte, eosinophil, mast cell, and neutrophil. The results are presented as a heatmap.

Additionally, the correlation between *SOWAHA* (including *SOWAHA* mutations and methylation) and major histocompatibility complex (MHC), immunomodulators, chemokines, and chemokine receptors genes were visualized as heatmaps using TISIDB.

**2.8 Association analysis between *SOWAHA* with tumor heterogeneity and stemness**

To calculate the DNA methylation based Stemness Scores (DNAss) and RNA based Stemness Score (RNAss) of *SOWAHA* in tumors, we utilized sangerbox 'Tumor stemness and gene expression' module, resulting in the final dataset comprising 37 types of cancer. Pearson correlation was then tested for each tumor, and the results were visualized using a lollipop chart. To investigate the link between *SOWAHA* expression and with tumor mutation burden (TMB)[22], Microsatellite Instability (MSI)[23], tumor neoantigen (NEO)[24], and Homologous Recombination Deficiency (HRD)[25], we utilized sangerbox 'Genomic heterogeneity and gene expression analysis' Module. Pearson's correlation was applied, and the results were visualized using a lollipop chart.

**2.9 Cell lines and cultures**

The colorectal cancer SW620 cell line was obtained from Tsinghua University. The cells were cultured in DMEM medium (Gibco, ThermoFisher, China) supplemented with 100 U/ml penicillin-streptomycin (HyClone, Utah, USA) and 10% fetal bovine serum (FBS, ExCell, China) at 37°C in a 5% $CO_2$ incubator.

**2.10 Cell Counting Kit-8 (CCK-8 assay)**

To assess cell viability after knocking down *SOWAHA*, we conduct a CCK-8 assay 72 hours post-transfection. Begin by seeding 7000 cells in a 96-well plate and allow them to adhere for 24 hours. The CCK-8 experiment was performed 72 hours after SOWAHA knockdown. Add the CCK-8 reagent and incubate for 2 hours at 37°C. Finally, measure the absorbance at 450 nm using a microplate reader. Analyze the results to determine the impact of the gene knockdown on cell viability.

**2.11 siRNA reverse transfection assay**

*SOWAHA*-siRNA (si*SOWAHA*) and NC-siRNA (siNC) were purchased from GenePharma (Shanghai, China), and mixed with transfection reagent (DharmaFECT1, Dharmacon Reagents, USA) and allowed to stand for 30 minutes. The resulting mixture was evenly distributed across the bottom of a 10 cm plate, into which $7 \times 10^5$ cells were seeded. After 72 hours of transfection, total mRNA was extracted to assess the transfection efficiency of the siRNA. The siRNA sequences of *SOWAHA* siRNA and negative control are as follows:

    *SOWAHA* siRNA-1: AGAAAUCUCUCGUCGACCUTT;
    *SOWAHA* siRNA-2: CUGCCAACAGAGCCACAGGTT;
    *SOWAHA* siRNA-3: GAAACCCACUUCGACGGUCTT;
    Negative control siRNA: UUCUCCGAACGUGUCACGUTT.

**2.12 Real-time fluorescence quantitative PCR (q-PCR)**

RNA extraction was performed using Trizol reagent (Invitrogen, USA), ensuring high-quality RNA with minimal contamination. The extracted RNA was then reverse transcribed into complementary DNA (cDNA) using the HiScript II Q RT SuperMix for q-PCR kit (Vazyme, China). Quantitative PCR (q-PCR) was conducted using real-time fluorescence to measure gene expression levels using ChamQ Universal SYBR q-PCR Master Mix (Vazyme, China), with β-actin serving as the reference gene. Data analysis was conducted using the $2^{-\Delta\Delta CT}$ method. The forward and reverse primers used are as follows:

*SOWAHA*-Forward primer 5'-3': CTTCAAGCAGTTCGTCAACAAC;
*SOWAHA*-Reverse primer 5'-3': AAGACCGTCGAAGTGGGTTTC.
β-actin-Forward primer 5'-3': CATGTACGTTGCTATCCAGGC;
β-actin -Reverse primer 5'-3': CTCCTTAATGTCACGCACGAT.

**2.13 RNA-seq experiments**

RNA-seq was performed on SW620 cell samples treated with siNC and si*SOWAHA*. RNA was extracted using Trizol reagent (Invitrogen, Thermo Fisher Scientific). Evaluate the quality and purity of the RNA using devices like the Bioanalyzer or Nanodrop, ensuring an RIN value above 7.0. Reverse transcription high-quality RNA into complementary DNA (cDNA) using a reverse transcriptase enzyme. Fragment the cDNA and add adapter sequences to create a library suitable for high-throughput sequencing. The sequencing library was then subjected to transcriptome sequencing on an Illumina Novaseq™ 6000 platform (LC-Bio Technologies (Hangzhou) Co., Ltd, China), with a minimum depth of 20,000 reads per cell. FASTQ reads from Illumina sequencing were mapped to the human genome reference GRCh38 using HiSat2[26]. Gene expression changes were analyzed with DESeq2, and differential expression was determined using cut-offs of $p < 0.05$ and $|\log2$ fold-change (log2FC)$| > 1$.

**2.14 Gene enrichment analysis**

DAVID (https://david.ncifcrf.gov/)[27] was used to perform Gene Ontology (GO) annotation, Disease Ontology (DO), and Kyoto Encyclopedia of Genes and Genomes (KEGG), extracting the top 20 or 30 terms for GO, DO, and KEGG. The results were visualized using the 'ggplot2' package in R. Gene set enrichment analysis (GSEA) was conducted via 'clusterProfiler' package. Besides, the 'cytoHubba' module of Cytoscape was used to identify the top 10 hub genes. Besides, the different alternative splicing (AS) genes were conducted KEGG and GSEA analysis using DAVID and visualized using R.

**2.15 Transcription factor (TF) binding sites prediction**

RNA-seq results revealed four differentially expressed transcription factors (log2FC > 1). Using NCBI (https://www.ncbi.nlm.nih.gov/), we identified the promoter sequence of *SOWAHA* in the 1.91 kb region from base 132,816,876 to 132,818,786. We then applied the 'scan' module in JASPAR (https://jaspar.genereg.net/) with a relative profile score threshold of 90% to identify transcription factor binding motifs within the *SOWAHA* promoter.

**2.16 Statistical analysis**

All data in this study was calculated, analyzed, and graphed using R software and GraphPad Prism 10. Gene expression differences are compared using either the Wilcoxon test or one-way ANOVA. The correlation between two groups is calculated using Pearson or Spearman correlation analysis. The log-rank test is employed to calculate significant differences in prognosis analysis. Student's t-test was used for the statistical analysis of q-PCR data. Significance levels are defined as follows: $p < 0.05$ (*), $p < 0.01$ (**), $p < 0.001$ (***), and $p < 0.0001$ (****).

## 3. Results

**3.1 *SOWAHA* is significantly low expressed in 12 types of cancers and high expressed in 14 types of cancers**

The mRNA expression of *SOWAHA* shows significant differences in various cancers. Specifically, *SOWAHA* was significantly low expressed in 12 types of cancer, including GBM, GBMLGG, LGG, KIRP, KIPAN, PRAD, KIRP, WT, SKCM, READ, TGCT, and KICH. *SOWAHA* was highly expressed in 14 types of cancer, including BRCA, CESC, ESCA, STES, COAD, COADREAD, STAD, LUSC, LIHC, THCA, OV, PAAD, UC, and PCPG (Figure 1A). In addition, the expression of *SOWAHA* across different TNM stages in 26 types of cancers and normal tissues was investigated. The results revealed significant differences in *SOWAHA* expression in 9 cancer types: LUAD, COAD, COADREAD, BRCA, ESCA, KIRP, KIPAN, HNSC, and PAAD (Figure 1B). Notably, these cancers are all of epithelial origin. Immunohistochemical images of *SOWAHA* in various cancers and normal tissues were retrieved from the HPA database. The protein expression of *SOWAHA* protein is lower in tumor than in normal tissue in almost all cancer types, such as KIRC(Figure 1C), GBM(Figure 1D), LGG (Figure 1D), COAD (Figure 1E), LIHC (Figure 1F), OV (Figure 1G), THCA (Figure 1H), and BRCA (Figure S1B), PRAD (Figure S1CC), TGCT (Figure S1D), CESC(Figure S1E), STAD(Figure S1FF),UCEC(Figure S1G), LUAD(Figure S1H), BLCA(Figure S1I), HNSC(Figure S1J), SKCM(Figure S1K), DLBC(Figure S1L), THCA(Figure S1M).

According to The Human Protein Atlas, the mRNA of *SOWAHA* is predominantly expressed in the brain, mainly in neuron, with a Tau specificity score of 0.84, indicating a high level of expression specificity in this tissue. However, it is also distributed across various other tissues, with cell type enrichment observed in colon enterocytes, respiratory ciliated cells in the lung, and thyroid glandular cells.

**3.2 The expression of *SOWAHA* has significant correlation with the prognosis of cancer patients**

A Cox hazards regression model was used to assess the association between *SOWAHA* gene expression and OS, DSS, and PFI with each tumor type. High expression of *SOWAHA* exhibited a notably poor prognosis with ACC and MESO, and low expression *SOWAHA* was notably associated with a poor prognosis of patients in KIPAN, GBMLGG, PAAD, UVM, KIRP, READ, HNSC, THYM, and THCA (Figure 2A). Kaplan-Meier survival plots showed significant correlation between high expression of *SOWAHA* and poor prognosis in ACC and MESO (Figure 2B-C). A significant correlation was observed between low *SOWAHA* expression and poor prognosis in KIPAN, GBMLGG, PAAD, HNSC, THCA, and READ (Figure 2D-I).

In addition, high expression of *SOWAHA* is significantly correlated with an unfavorable prognosis of DSS in 2 types of cancer patients: ACC, and MESO, and low expression of *SOWAHA* is significantly correlated with unfavorable prognosis of DSS in 9 tumor types: KIPAN, KIRP, GBMLGG, THYM, PAAD, UVM, HNSC, BRCA, and COADREAD (Figure S2A). High expression of *SOWAHA* is significantly correlated with an unfavorable prognosis of PFI in ACC, and low expression of *SOWAHA* is significantly correlated with unfavorable prognosis of PFI in PAAD, KIRP, GBMLGG, PCPG, BRCA, SKCM, SKCM-M, HNSC, and UVM (Figure S2B). Interestingly, all cancer types that show a significant correlation between low *SOWAHA* expression and poor prognosis are epithelial in origin, including GBMLGG, which arise from neuroepithelial cells in embryology.

**3.3 *SOWAHA* displays mutations across 23 different cancer types**

The predominant forms of mutations in *SOWAHA* were missense (53%) and synonymous (32%) mutations (Figure 3A), with KIRC (5.28%), UCS (5.26%), and CHOL (2.78%) being the most frequently affected cancer types (Figure 3B). The CNA types of *SOWAHA* are significantly different in GBMLGG, LGG, LUAD, BRCA, KIPAN, STAD, PAAD, OV, PCPG, and UCS (Figure 3C).

**3.4 *SOWAHA* is low expressed in multiple cancer driver genes mutation types of cancer**

*SOWAHA* is under expressed in EGFR-mutated COAD and LUSC, but overexpressed in EGFR-mutated GBM (Figure 4A-C). In PIK3CA-mutated cancers, *SOWAHA* shows low expression in COAD but elevated levels in BRCA, including BRCA-LumA and BRCA-LumB subtypes (Figure 4D-G). *SOWAHA* expression is reduced in MLH1-mutated COAD, BRCA-LumB, and SKCM (Figure 4H-J), as well as in MSH2-mutated COAD and LUSC (Figure 4K-L). Additionally, *SOWAHA* is downregulated in PMS2, MSH4, and MSH6-mutated COAD (Figure 4M-O). Interestingly, *SOWAHA* is overexpressed in TP53-mutated BRCA-LumB, COAD, STAD, and UCEC, but downregulated in BRCA, HNSC, HNSC+HPV, GBM, and LGG (Figure S3A-I). Lastly, *SOWAHA* exhibits lower expression in KRAS-mutated COAD, PAAD, TGCT, UCEC, BLCA, and STAD (Figure S3J-O).

**3.5 *SOWAHA* methylation at different sites correlates variably with cancer patient survival**

We analyzed *SOWAHA* DNA methylation levels in tumor and normal tissues, revealing significant differences in *SOWAHA* methylation in pan-cancer. The methylation levels of *SOWAHA* promoter were significantly increased in BLCA, BRCA, CESC, COAD, ESCA, HNSC, KIRC, KIRP, LIHC, LUAD, LUSC, PRAD, SKCM, THCA, and UCEC patient tissues. (Figure 5A).

The 1st exon represents a crucial region for gene transcription. When this site undergoes heavy methylation, it can obstruct transcription initiation, potentially altering gene expression, particularly in diseases like cancer. There is significant negative correlation between 1stExon-Island-cg02699218 methylation of *SOWAHA* and the survival probability of ACC, MESO, READ (Figure 5B-D), and UCEC, SKCM, and SARC (Figure S4G-I), but is significant positive correlation with BRCA, and UVM (Figure S4J-K). There is significant negative correlation between 1stExon-island-cg01435564 and 1stExon-island-cg04044664 methylation of *SOWAHA* and survival probability of ACC, MESO, and READ (Figure 5E-J). Besides, there is significant positive correlation between 1stExon-island-cg01435564 methylation of *SOWAHA* and survival probability of UVM and BLCA (Figure S4A-B). There is significant negative correlation between 1stExon-island-cg04044664 methylation of *SOWAHA* and survival probability of COAD, PAAD, and UCEC (Figure S4C-E), while opposite in STAD (FigureS4F). In addition, there is significant negative correlation between 1stExon-island-cg26399201 methylation of *SOWAHA* and survival probability of UCEC (FigureS4L).

Methylation in the TSS200 region is typically associated with gene silencing. Elevated methylation levels in this area can inhibit the binding of transcription factors and other regulatory proteins, thus reducing gene expression. We observed a significant positive correlation between TSS200-Island-cg26066519 methylation of *SOWAHA* and survival probability in BRCA and LGG (Figure S5A-B), while an opposite trend was noted in BLCA, MESO, and SARC (Figure S5C-E). Additionally, there was a significant negative correlation between TSS200-Island-cg12923585 methylation of *SOWAHA* and survival probability in BLCA, UCEC, LGG, and UVM (Figure S5F-I), whereas the trend was reversed in ACC, GBM, and LAML (Figure S5J-L). Lastly, TSS200-Island-cg18840461 methylation of *SOWAHA* showed a significant negative association with survival probability in PAAD and UVM (Figure S5M-N), while the contrary was observed in SKCM (Figure S5O).

### 3.6 *SOWAHA* expression shows significant correlation with Immune Infiltration Score across cancer types

The immune infiltration score reflects the impact of the immune system on the tumor microenvironment by calculating the proportion or quantity of different types of immune cells in tumor or tissue samples. We utilized ESTIMATE to calculate the StromalScore, ImunneScore, and ESTIMATEScore, and found that the expression of *SOWAHA* was significantly correlated with immune infiltration in 26 cancer species, of which 4 were significantly positively correlated in ESTIMATEScore, namely NB, LIHC, HNSC, and BRCA. There were 15 significant negative correlations, namely ALL-R, ACC, PCPG, TGCT, WT, BLCA, STAD, COADREAD, COAD, KIPAN, KIRP, ESCA, CESC, LGG, and GBMLGG. and SKCM-P (Figure 6A). Interestingly, *SOWAHA* is positively correlation with StromalScore, ImunneScore, and ESTIMATEScore in GBM, but is significantly negative correlation with StromalScore, ImunneScore, and ESTIMATEScore in LGG and GBMLGG, which suggests *SOWAHA* has distinct biological functions across different stages or types of gliomas. The detail of the correlation between *SOWAHA* expression and StromalScore, ImmuneScore and ESTIMATEScore for GBMLGG, KIPAN, and ACC is shown in Figure 6B-J.

### 3.7 Significantly association between *SOWAHA* expression, CNA, methylation and TIIC across 30 types of cancer

To explore the function of *SOWAHA* in the TME, we assessed the relationship between *SOWAHA* expression and TIIC. The results showed that *SOWAHA* is negatively associated with infiltration of $CD8^+$T cells, $CD4^+$ T cells, Tfh, Tgd, Treg, macrophages, eosinophils, Th1, Th2, B cells, NK cells, mast cells, monocytes, and neutropils in the majority cancer types (Figure 7A). Furthermore, *SOWAHA* is positively associated with the Th17 in many types of cancer, especially in ESCA. Besides, the correlation between CNA of *SOWAHA* and TIIC is generally positive compared to the correlation observed between *SOWAHA* and TIIC, and CNA of *SOWAHA* is positively associated with all immune cells except MemB in ESCA, GBM, HNSC, LUSC, LUAD, STAD, TGCT, and UVM, while the correlation between CNA of *SOWAHA* and TIIC is negative in ACC (Figure 7B). While methylation of *SOWAHA* is strong positively associated with almost all immune cells in ACC, BLCA, CESC, CHOL, and LUSC, but is opposite in KIRP, LGG, and PRAD (Figure 7C).

### 3.8 *SOWAHA* is negative correlation with multiple MHC-related genes

Cancer cells evade immune detection by downregulating MHC molecule expression or modifying antigen presentation mechanisms to facilitate immune escape (35). To investigate the link between *SOWAHA* and MHC genes, we analyzed the correlation between *SOWAHA* and MHC-related genes. Our results revealed that *SOWAHA* expression is negatively correlated with all MHC-related genes in ACC, BLCA, KIRP, LGG, PAAD, TGCT, and UVM, but shows the opposite trend in KICH and UCS (Figure 7D). Next, we assessed the relationship between the CNA of *SOWAHA* and MHC genes, finding that CNA of *SOWAHA* is positively associated with nearly all MHC genes across most cancer types, except in ACC and KIRC (Figure 7E). Additionally, *SOWAHA* methylation levels are positively correlated with MHC-related genes in ACC, BLCA, CESC, COAD, LUAD, LUSC, READ, SKCM, STAD, TGCT, UCEC, UCS, and UVM, but display an inverse correlation in KICH, LGG, and PRAD (Figure 7F).

### 3.9 *SOWAHA* is negative relation with many chemokines and receptors

We investigated the relationship between *SOWAHA* and chemokines, as well as their receptors, in pan-cancer. The results revealed that *SOWAHA* is positively correlated with CCL15 and CCL28 across various cancer types, but negatively correlated with many other chemokines in multiple cancers, except in ESCA and TGCT (Figure S6A). The CNA of *SOWAHA* is positively correlated with nearly all chemokines in LUSC and TGCT, while a weak correlation was observed between the CNA of *SOWAHA* and CCL1, CCL16, CCL24, CCL25, and CCL26 in most cancers (Figure S6B). The methylation of *SOWAHA* shows a minimal correlation with CCL1 in nearly all cancer types but is positively correlated with CCL2, CCL3, CCL4, CCL5, CCL8, CCL13, CCL14, CXCL9, CXCL10, CXCL11, CXCL12, XCL1, and XCL2 in ACC, BLCA, CESC, CHOL, and COAD (Figure S6C). *SOWAHA* shows weak correlation with CCR3, CCR9, and XCR1 in most cancers, but is positively correlated with CCR6 in the majority of cancers, except ACC (Figure S6D). The CNA of *SOWAHA* is weakly correlated with CCR3 and CCR9 in most cancers but shows a positive correlation with almost all chemokine receptors in HNSC, LUSC, and STAD (Figure S6E). Furthermore, *SOWAHA* methylation is weakly related to CCR3, CCR9, and XCR1 in most cancers, yet significantly positively correlated with most chemokine receptors in ACC, BLCA, CESC, CHOL, and LUAD. Additionally, *SOWAHA* methylation is strongly positively correlated with

CCR6 and CCR8 in OV (Figure S6F).

**3.10 The expression of *SOWAHA* is significantly associated with tumor stemness**
We observed there is significant correlation between *SOWAHA* and the DNA methylation based Stemness Scores (DNAss)[28] in 12 types of tumors, with 5 showing significant positive correlations, such as LUAD, STAD, PRAD, THCA, and PCPG, and 7 showing significant negative correlations, such as GBMLGG, CESC, BRCA, KIRP, KIPAN, PAAD, and TGCT (Figure 8A). There is significant correlation between *SOWAHA* and the RNA based Stemness Scores (RNAss)[29] 17 tumor types, with 7 showing significant positive correlations, such as GBMLGG, LGG, ESCA, KIPAN, LIHC, PCPG, ACC, while 10 showing significant negative correlations, including LUAD, BRCA, SARC, PRAD, THYM, THCA, MESO, TGCT, SKCM, and UVM (Figure 8B).

**3.11 The expression of *SOWAHA* is significantly associated with tumor heterogeneity**
We observed the expression of *SOWAHA* is significantly associated with TMB in 8 types of tumors, with 2 showing significant positive correlations, such as ESCA and CHOL, and 6 showing significant negative correlations, such as CESC, COAD, COADREAD, PRAD, LUSC, and READ (Figure 8C). The expression of *SOWAHA* is significantly associated with MSI in 10 types of tumors, with 4 showing significant positive correlations, such as GBMLGG, KIPAN, TGCT, and SKCM, and 6 showing significant negative correlations, such as COAD, COADREAD, BRCA, STAD, PRAD, and HNSC (Figure 8D). The expression of *SOWAHA* is significantly associated with HRD in 14 types of tumors, with 8 significant positive correlations, such as COAD, STES, STAD, UCEC, HNSC, LIHC, BLCA, and ACC, and 6 significant negative correlations, such as GBM, GBMLGG, LGG, BRCA, PAAD, and KICH (Figure 8E). Significant correlations were observed between *SOWAHA* expression and NEO in 6 types of tumors, with 1 showing a significant positive correlation, CHOL, and 5 showing significant negative correlations, such as GBM, COAD, COADREAD, BRCA, and HNSC (Figure 8F).

**3.12 *SOWAHA* positively correlates with many immune inhibitors and stimulators**
We investigated the correlation between *SOWAHA* expression and immunoinhibitors and observed that *SOWAHA* negatively correlates with many inhibitors in multiple cancer types, but is significantly positive corelated with PVRL2 in ESCA, and significantly positive correlated with TGFBR1 in THCA. But *SOWAHA* expression is low corelated with KIR2DL1 and KIR2DL3 in almost all cancer types (Figure S7A). The same situation is also observed in the correlation of CNA of *SOWAHA,* methylation *of SOWAHA* and immunoinhibitors (Figure S7B). The CNA of *SOWAHA* is positively corelates with many inhibitors in multiple cancer types. The methylation of *SOWAHA* is positively related to many immunomodulators in majority cancer types, except LGG and PRAD (Figure S7C).
We investigated the correlation between *SOWAHA* expression and immunostimulators and observed that *SOWAHA* is significantly positive related to HHLA2 and TNFRSF14, TNFSF13, TNFSF15, in ESCA. But *SOWAHA* is minimal corelated with all immunostimulators in MESO ((Figure S7D).
The CNA of *SOWAHA* is positively correlated with many immunositimulators, especially in LUSC, and STAD ((Figure S7E). The methylation of *SOWAHA* is strong positive correlated with almost all immunostimulators in ACC, BLCA, CESC, CHOL, COAD, LUAD, and is strong positive association with TNFRSF8, and TNFRSF9 in OV (Figure S7F).

### 3.13 Function analysis revealed that *SOWAHA* is significantly connected with metabolic pathways in carcinoma

We knocked down *SOWAHA* in the SW620 cell line using siRNA, achieving a knockdown efficiency of over 90% (Figure 9A). RNA-seq was then performed on si*SOWAHA* samples along with siNC samples. Compared to siNC, *SOWAHA* knockdown resulted in 54 significantly upregulated genes and 70 significantly downregulated genes (Figure 9B-C). KEGG pathway enrichment analysis revealed significant enrichment in pathways: 'Cytokine-cytokine receptor interaction', 'PI3K-AKT signaling pathway', 'Pathways in cancer', 'NF-kappa B signaling pathway'. Besides, 'Breast cancer', 'Small cell lung cancer', 'Melanoma' are also significantly enriched. (Figure 9D).

GO enrichment analysis revealed that DEGs were primarily enriched in 'pallium development', 'Trans-1,2-dehydrobenzene-1,2-diol dehydrogenase activity', 'indanol dehydrogenase activity', 'oxidoreductase activity, acting on NAD(P)H, quinone' and many metabolic pathways. Additionally, 'T cell migration' and the 'fibroblast apoptosis process' were also enriched (Figure10A). DO enrichment analysis results indicated that 'gastrointestinal lymphoma,' 'gastrointestinal system cancer,' and 'congenital nervous system abnormalities,' among others, were significantly enriched (Figure 10B). DO-based GSEA analysis revealed that significantly upregulation in 'hereditary Wilms' tumor', 'acute leukemia', 'pancreas disease', 'sarcoma', 'multiple myeloma', and downregulation in 'pancreatic ductal adenocarcinoma', and 'colonic disease', among others (Figure 10C).

GSEA analysis revealed significantly upregulation in 'Glycosylphosphatidylinositol (GPI) Anchor Biosynthesis', 'P53 signaling pathway', 'Ubiquitin-Mediated Proteolysis', 'RNA Degradation', 'Sulfur Metabolism', 'Nucleotide Excision Repair', and 'Basal Transcription factors' (Figure 10D). In addition, the 10 hub genes are respectfully FGF8, FGF18, CXCL8, CSF3, NT5E, TGFB2, CCND1, CXCL1, CX3CL1, and APOE (Figure10G).

### 3.14 TBX4 and FOXP2 may bind to the promoter of *SOWAHA*

By RNA-seq, we identified four transcription factors with differential expression (log2FC > 1): FOXP2, ZFP91-CNTF, PAX6, and TBX4 (Figure 9E). We used JASPAR to predict their binding motifs for *SOWAHA* and found that TBX4 and FOXP2 can bind to the promoter of *SOWAHA* (Table 2, Figure 9G). FOXP2 is upregulated in si*SOWAHA* samples, while TBX4 is downregulated in these samples. Additionally, FOXP2 underwent a Skipped Exon (SE) event compared to siNC.

### 3.15 KEGG analysis of AS indicated *SOWAHA* is connected with carcinoma

The types of alternative splicing (AS) observed in the differentially expressed samples, including Alternative 3' Splice Site (A3S), Alternative 5' Splice Site (A5S), Mutually Exclusive Exons (MXE), Retained Intron (RI), and Skipped Exon (SE), are shown in Figure 9F. Among these, SE and MXE are the predominant AS types, having the greatest impact on gene function. KEGG analysis of AS genes indicated significant terms are 'Viral carcinogenesis', 'autophagy-animal', 'RNA Transport', 'EGFR tyrosine kinase inhibitor resistance', 'Pyrimidine metabolism', etc. (Figure 10E). KEGG-based GSEA of all AS genes were primarily significant enriched in 'P53 signaling pathway', 'Cell cycle', 'RNA Degradation', 'sulfur metabolism', 'basal transcription factors' (Figure10F).

### 3.16 The enhanced proliferation of SW620 cells after *SOWAH* knockdown

Under the microscope, SW620 cells with *SOWAHA* knockdown appeared more densely packed, with a more spread-out and uniform morphology, suggesting enhanced growth and proliferation (Figure 10H). CCK-8 assay results showed that the cells proliferation and viability in si*SOWAHA* groups is better than siNC groups (Figure 10I).

**Discussion**

The mRNA expression of *SOWAHA* is significantly different expressed in 26 types of cancer. Bu the protein expression of *SOWAHA* in these cancer types is lower than the normal tissue. Low expression of *SOWAHA* is significantly correlated with unfavorable prognosis of OS, DSS, and PFI in epithelial cancer types. The low expression of *SOWAHA* is associated with cancer progression, suggesting that it may have a role in suppressing cancer progression under normal conditions. Besides, knockdown of *SOWAHA* in SW620 cells exhibited better growth, these indicates that *SOWAHA* is a cancer suppressor gene in COAD.

The significant discrepancy between mRNA and protein expression levels of *SOWAHA* in SW620 cell line may be influenced by factors such as translation regulation and epigenetic modulation. GO analysis results indicated *SOWAHA* is significantly related with activity of many enzymes and metabolism processes, and the results of KEGG and GSEA analysis showed that *SOWAHA* is related with many signaling pathways in cancer. DO analysis results showed it is significantly enriched in 'gastrointestinal lymphoma', 'gastrointestinal system cancer', and 'congenital nervous system abnormality', etc. Knockdown of *SOWAHA* resulted in the different expression of many AS genes, which is associated with multiple cancer-related pathways, such as 'Viral carcinogenesis', 'autophagy-animal', 'EGFR tyrosine kinase inhibitor resistance', etc. GSEA enrichment analyses of these genes revealed significant enrichment in various cancer-related metabolic reprogramming pathways, such as GPI anchor biosynthesis, Cell cycle, P53 pathway, PI3K-AKT pathway, RNA degradation, Ubiquitin-mediated proteolysis, Basal transcription factors, and Nucleotide excision repair.

In cancer cells, abnormal expression or activity changes of certain transcription factors may affect the binding of splicing factors and the selection of splicing sites, potentially leading to alterations in splicing patterns[30].Our research found 2 potential transcription factors of *SOWAHA*, T-box transcription factor-4 (TBX4), and Forkhead Box P2 (FOXP2), both are biomarker in some cancer types. TBX4, a member of the T-box transcription factor family, plays a crucial role in both embryonic development and cancer[31]. During embryonic development, TBX4 is essential for the proper formation of lower limbs and the development of the lungs[32]. The expression of TBX4 can become dysregulated in cancer. Abnormal levels of TBX4 have been observed in various cancers, including LUSC, BRCA, and CHOL[33-35]. In these cancers, TBX4 may contribute to tumor progression by affecting cell proliferation and invasion[36].

FOXP2 is crucial for the development of the nervous system during embryogenesis[37]. It regulates the proliferation, migration, and differentiation of neurons, influencing the formation of the brain and spinal cord[38]. Its abnormal expression may affect normal neural development, potentially indirectly promoting the progression of neurodevelopmental cancers[39]. It has been reported that

*SOWAHA* is involved in embryonic development in chicken embryo[40]. Our research found *SOWAHA* is significantly corelated with pallium development, and transcriptional misregulation in cancer, and glioma, COAD, BRCA, LUSC, STAD, etc. Some studies suggest that FOXP2 may play a role in the progression of GBM[39, 41], COAD[33], BRCA[42], LUSC[43], STAD[44], etc., but the specific mechanisms and functions are still under investigation. Our findings provide a reference for this issue.

*SOWAHA* is expressed at low levels in several cancer types associated with driver mutations, particularly in COAD. DNA methylation of *SOWAHA* shows positive correlations with TIIC and MHC, and is significantly associated with poor prognosis across various cancers. Additionally, *SOWAHA* expression is notably linked with tumor heterogeneity and stemness in multiple cancer types. Besides, *SOWAHA* is significantly associated with DNAss, RNAss, TMB, MSI, HRD, and NEO in various cancer types, which originate from epithelial tissues. Importantly, the function enrichment results showed *SOWAHA* is significantly related with many pathways in cancer. These findings suggest that *SOWAHA* plays a crucial role in cancer occurrence and progression and may serve as a valuable diagnostic and prognostic biomarker in some cancer types, including COAD.

This study has some limitations, while it systematically explores the correlation between *SOWAHA* and various cancer types and identifies potential biological processes and signaling pathways involved in its regulation, it largely relies on bioinformatics methods. Further experimental validation is needed to confirm these findings. Additionally, the biological functions and molecular mechanisms of *SOWAHA* in tumors, particularly within the TIME, require further investigation.

**Author Statement**
Xiaohong Yi, Hua Zhong, Yumei Wang, contributed to the conception of the study. Data collection, analysis, and figure generation were performed by Xiaohong Yi, Claire H. Zhao, and Lijun Huang, purchase of reagents and consumables was completed by Yuhui Chen. Experiments were conducted by Xiaohong Yi, Xianwen Zhang. The manuscript was drafted by Xiaohong Yi, and revised by Hua Zhong, Yumei Wang. All authors contributed to the article and approved the final submitted version.

**Ethics approval and consent to participate**
Not applicable.

**Declaration of Competing Interest**
The authors declare that they have no financial conflicts of interest related to this paper.

**Funds**
This work was funded by the Xinglin Scholars Program of Chengdu University of Traditional Chinese Medicine (No.330023317), and Research Project of Sichuan Traditional Chinese Medicine Administration (No.2024MS170).

**Acknowledgements**
We would like to thank Prof. Dong Wang (School of Basic Medical Sciences, CDUTCM, Chengdu) for his invaluable assistance with this article. We are also grateful to all members of our laboratory for their support in this research. Finally, we extend our gratitude to the patients, physicians, and scientists involved in TCGA, NCBI, HPA, and other databases.**References**

[1] F. Bray, M. Laversanne, H. Sung, J. Ferlay, R.L. Siegel, I. Soerjomataram, A. Jemal, Global cancer statistics 2022: GLOBOCAN estimates of incidence and mortality worldwide for 36 cancers in 185 countries, CA Cancer J Clin, 74 (2024) 229-263.

[2] S.T. Crooke, J.L. Witztum, C.F. Bennett, B.F. Baker, RNA-Targeted Therapeutics, Cell Metab, 27 (2018) 714-739.

[3] X. Zhang, L. Zhu, H. Zhang, S. Chen, Y. Xiao, CAR-T Cell Therapy in Hematological Malignancies: Current Opportunities and Challenges, Front Immunol, 13 (2022) 927153.

[4] N. Chatterjee, T.G. Bivona, Polytherapy and Targeted Cancer Drug Resistance, Trends Cancer, 5 (2019) 170-182.

[5] J. Kim, J. Lim, S.W. Lee, J.Y. Park, D.S. Suh, J.H. Kim, Y.M. Kim, D.Y. Kim, Immune checkpoint inhibitors with chemotherapy for primary advanced mismatch repair-deficient endometrial cancer: A cost-effectiveness analysis, Gynecol Oncol, 179 (2023) 106-114.

[6] X. Cai, Y. Zhang, Molecular evolution of the ankyrin gene family, Mol Biol Evol, 23 (2006) 550-558.

[7] Y. Yin, S. Yang, Z. Huang, Z. Yang, C. Zhang, Y. He, RNA methylation-related genes INHBB and SOWAHA are associated with MSI status in colorectal cancer patients and may serve as prognostic markers for predicting immunotherapy efficacy, Carcinogenesis, 45 (2024) 337-350.

[8] Y. Mori, H. Yokota, I. Hoshino, Y. Iwatate, K. Wakamatsu, T. Uno, H. Suyari, Deep learning-based gene selection in comprehensive gene analysis in pancreatic cancer, Sci Rep, 11 (2021) 16521.

[9] S. Lv, Z. Qian, J. Li, S. Piao, J. Li, Identification and Validation of a Hypoxia-Immune-Based Prognostic mRNA Signature for Oral Squamous Cell Carcinoma, J Oncol, 2022 (2022) 5286251.

[10] L.R. Nassar, G.P. Barber, A. Benet-Pagès, J. Casper, H. Clawson, M. Diekhans, C. Fischer, J.N. Gonzalez, A.S. Hinrichs, B.T. Lee, C.M. Lee, P. Muthuraman, B. Nguy, T. Pereira, P. Nejad, G. Perez, B.J. Raney, D. Schmelter, M.L. Speir, B.D. Wick, A.S. Zweig, D. Haussler, R.M. Kuhn, M. Haeussler, W.J. Kent, The UCSC Genome Browser database: 2023 update, Nucleic Acids Res, 51 (2023) D1188-D1195.

[11] A. Colaprico, T.C. Silva, C. Olsen, L. Garofano, C. Cava, D. Garolini, T.S. Sabedot, T.M. Malta, S.M. Pagnotta, I. Castiglioni, M. Ceccarelli, G. Bontempi, H. Noushmehr, TCGAbiolinks: an R/Bioconductor package for integrative analysis of TCGA data, Nucleic Acids Res, 44 (2016) e71.

[12] The Genotype-Tissue Expression (GTEx) project, Nat Genet, 45 (2013) 580-585.

[13] E. Cerami, J. Gao, U. Dogrusoz, B.E. Gross, S.O. Sumer, B.A. Aksoy, A. Jacobsen, C.J. Byrne, M.L. Heuer, E. Larsson, Y. Antipin, B. Reva, A.P. Goldberg, C. Sander, N. Schultz, The cBio cancer genomics portal: an open platform for exploring multidimensional cancer genomics data, Cancer Discov, 2 (2012) 401-404.

[14] F. Pontén, M. Gry, L. Fagerberg, E. Lundberg, A. Asplund, L. Berglund, P. Oksvold, E. Björling, S. Hober, C. Kampf, S. Navani, P. Nilsson, J. Ottosson, A. Persson, H. Wernérus, K. Wester, M. Uhlén, A global view of protein expression in human cells, tissues, and organs, Mol Syst Biol, 5 (2009) 337.

[15] M. Uhlén, L. Fagerberg, B.M. Hallström, C. Lindskog, P. Oksvold, A. Mardinoglu, Å. Sivertsson, C. Kampf, E. Sjöstedt, A. Asplund, I. Olsson, K. Edlund, E. Lundberg, S. Navani, C.A. Szigyarto, J. Odeberg, D. Djureinovic, J.O. Takanen, S. Hober, T. Alm, P.H. Edqvist, H. Berling, H. Tegel, J. Mulder, J. Rockberg, P. Nilsson, J.M. Schwenk, M. Hamsten, K. von Feilitzen, M. Forsberg, L. Persson, F.

Johansson, M. Zwahlen, G. von Heijne, J. Nielsen, F. Pontén, Proteomics. Tissue-based map of the human proteome, Science, 347 (2015) 1260419.

[16] W. Shen, Z. Song, X. Zhong, M. Huang, D. Shen, P. Gao, X. Qian, M. Wang, X. He, T. Wang, S. Li, X. Song, Sangerbox: A comprehensive, interaction-friendly clinical bioinformatics analysis platform, Imeta, 1 (2022) e36.

[17] F. Martínez-Jiménez, F. Muiños, I. Sentís, J. Deu-Pons, I. Reyes-Salazar, C. Arnedo-Pac, L. Mularoni, O. Pich, J. Bonet, H. Kranas, A. Gonzalez-Perez, N. Lopez-Bigas, A compendium of mutational cancer driver genes, Nat Rev Cancer, 20 (2020) 555-572.

[18] T. Li, J. Fu, Z. Zeng, D. Cohen, J. Li, Q. Chen, B. Li, X.S. Liu, TIMER2.0 for analysis of tumor-infiltrating immune cells, Nucleic Acids Res, 48 (2020) W509-W514.

[19] Y. Li, D. Ge, C. Lu, The SMART App: an interactive web application for comprehensive DNA methylation analysis and visualization, Epigenetics Chromatin, 12 (2019) 71.

[20] V. Modhukur, T. Iljasenko, T. Metsalu, K. Lokk, T. Laisk-Podar, J. Vilo, MethSurv: a web tool to perform multivariable survival analysis using DNA methylation data, Epigenomics, 10 (2018) 277-288.

[21] B. Ru, C.N. Wong, Y. Tong, J.Y. Zhong, S.S.W. Zhong, W.C. Wu, K.C. Chu, C.Y. Wong, C.Y. Lau, I. Chen, N.W. Chan, J. Zhang, TISIDB: an integrated repository portal for tumor-immune system interactions, Bioinformatics, 35 (2019) 4200-4202.

[22] T.A. Chan, M. Yarchoan, E. Jaffee, C. Swanton, S.A. Quezada, A. Stenzinger, S. Peters, Development of tumor mutation burden as an immunotherapy biomarker: utility for the oncology clinic, Ann Oncol, 30 (2019) 44-56.

[23] M. Baretti, D.T. Le, DNA mismatch repair in cancer, Pharmacol Ther, 189 (2018) 45-62.

[24] A. Esprit, W. de Mey, R. Bahadur Shahi, K. Thielemans, L. Franceschini, K. Breckpot, Neo-Antigen mRNA Vaccines, Vaccines (Basel), 8 (2020).

[25] K.D. Doig, A.P. Fellowes, S.B. Fox, Homologous Recombination Repair Deficiency: An Overview for Pathologists, Mod Pathol, 36 (2023) 100049.

[26] D. Kim, J.M. Paggi, C. Park, C. Bennett, S.L. Salzberg, Graph-based genome alignment and genotyping with HISAT2 and HISAT-genotype, Nat Biotechnol, 37 (2019) 907-915.

[27] G. Dennis, Jr., B.T. Sherman, D.A. Hosack, J. Yang, W. Gao, H.C. Lane, R.A. Lempicki, DAVID: Database for Annotation, Visualization, and Integrated Discovery, Genome Biol, 4 (2003) P3.

[28] T.M. Malta, A. Sokolov, A.J. Gentles, T. Burzykowski, L. Poisson, J.N. Weinstein, B. Kamińska, J. Huelsken, L. Omberg, O. Gevaert, A. Colaprico, P. Czerwińska, S. Mazurek, L. Mishra, H. Heyn, A. Krasnitz, A.K. Godwin, A.J. Lazar, J.M. Stuart, K.A. Hoadley, P.W. Laird, H. Noushmehr, M. Wiznerowicz, Machine Learning Identifies Stemness Features Associated with Oncogenic Dedifferentiation, Cell, 173 (2018) 338-354 e315.

[29] A. Bagaev, N. Kotlov, K. Nomie, V. Svekolkin, A. Gafurov, O. Isaeva, N. Osokin, I. Kozlov, F. Frenkel, O. Gancharova, N. Almog, M. Tsiper, R. Ataullakhanov, N. Fowler, Conserved pan-cancer microenvironment subtypes predict response to immunotherapy, Cancer Cell, 39 (2021) 845-865 e847.

[30] X.D. Fu, M. Ares, Jr., Context-dependent control of alternative splicing by RNA-binding proteins, Nat Rev Genet, 15 (2014) 689-701.

[31] J.A. Karolak, C.L. Welch, C. Mosimann, K. Bzdęga, J.D. West, D. Montani, M. Eyries, M.P. Mullen, S.H. Abman, M. Prapa, S. Gräf, N.W. Morrell, A.R. Hemnes, F. Perros, R. Hamid, M.P.O. Logan, J. Whitsett, C. Galambos, P. Stankiewicz, W.K. Chung, E.D. Austin, Molecular Function and Contribution of TBX4 in Development and Disease, Am J Respir Crit Care Med, 207 (2023) 855-864.

[32] M. Logan, C.J. Tabin, Role of Pitx1 upstream of Tbx4 in specification of hindlimb identity, Science,

283 (1999) 1736-1739.

[33] L.E. Kelemen, X. Wang, Z.S. Fredericksen, V.S. Pankratz, P.D. Pharoah, S. Ahmed, A.M. Dunning, D.F. Easton, R.A. Vierkant, J.R. Cerhan, E.L. Goode, J.E. Olson, F.J. Couch, Genetic variation in the chromosome 17q23 amplicon and breast cancer risk, Cancer Epidemiol Biomarkers Prev, 18 (2009) 1864-1868.

[34] I.L. Lai, Y.S. Chang, W.L. Chan, Y.T. Lee, J.C. Yen, C.A. Yang, S.Y. Hung, J.G. Chang, Male-Specific Long Noncoding RNA TTTY15 Inhibits Non-Small Cell Lung Cancer Proliferation and Metastasis via TBX4, Int J Mol Sci, 20 (2019).

[35] M. Zong, L. Jia, L. Li, Expression of novel tumor markers of pancreatic adenocarcinomas in intrahepatic cholangiocarcinomas, Onco Targets Ther, 6 (2013) 19-23.

[36] T. Xie, J. Liang, N. Liu, C. Huan, Y. Zhang, W. Liu, M. Kumar, R. Xiao, J. D'Armiento, D. Metzger, P. Chambon, V.E. Papaioannou, B.R. Stripp, D. Jiang, P.W. Noble, Transcription factor TBX4 regulates myofibroblast accumulation and lung fibrosis, J Clin Invest, 126 (2016) 3063-3079.

[37] J. den Hoed, K. Devaraju, S.E. Fisher, Molecular networks of the FOXP2 transcription factor in the brain, EMBO Rep, 22 (2021) e52803.

[38] S.L. Hickey, S. Berto, G. Konopka, Chromatin Decondensation by FOXP2 Promotes Human Neuron Maturation and Expression of Neurodevelopmental Disease Genes, Cell Rep, 27 (2019) 1699-1711 e1699.

[39] J. Plata-Bello, H. Fariña-Jerónimo, I. Betancor, E. Salido, High Expression of FOXP2 Is Associated with Worse Prognosis in Glioblastoma, World Neurosurg, 150 (2021) e253-e278.

[40] R. Zhang, F. Yao, X. Cheng, M. Yang, Z. Ning, Identification of candidate genomic regions for egg yolk moisture content based on a genome-wide association study, BMC Genomics, 24 (2023) 110.

[41] Q. He, L. Zhao, Y. Liu, X. Liu, J. Zheng, H. Yu, H. Cai, J. Ma, L. Liu, P. Wang, Z. Li, Y. Xue, circ-SHKBP1 Regulates the Angiogenesis of U87 Glioma-Exposed Endothelial Cells through miR-544a/FOXP1 and miR-379/FOXP2 Pathways, Mol Ther Nucleic Acids, 10 (2018) 331-348.

[42] B.G. Cuiffo, A. Campagne, G.W. Bell, A. Lembo, F. Orso, E.C. Lien, M.K. Bhasin, M. Raimo, S.E. Hanson, A. Marusyk, D. El-Ashry, P. Hematti, K. Polyak, F. Mechta-Grigoriou, O. Mariani, S. Volinia, A. Vincent-Salomon, D. Taverna, A.E. Karnoub, MSC-regulated microRNAs converge on the transcription factor FOXP2 and promote breast cancer metastasis, Cell Stem Cell, 15 (2014) 762-774.

[43] W. Shu, M.M. Lu, Y. Zhang, P.W. Tucker, D. Zhou, E.E. Morrisey, Foxp2 and Foxp1 cooperatively regulate lung and esophagus development, Development, 134 (2007) 1991-2000.

[44] P. Xu, X. Zhang, J. Cao, J. Yang, Z. Chen, W. Wang, S. Wang, L. Zhang, L. Xie, L. Fang, Y. Xia, Z. Xuan, J. Lv, H. Xu, Z. Xu, The novel role of circular RNA ST3GAL6 on blocking gastric cancer malignant behaviours through autophagy regulated by the FOXP2/MET/mTOR axis, Clin Transl Med, 12 (2022) e707.

Table 1. Full names and abbreviations of the 45 cancer types in TCGA.

| Abbreviation | Cancer types |
|---|---|
| ACC | Adrenocortical carcinoma |
| BLCA | Bladder Urothelial Carcinoma |
| BRCA | Breast invasive carcinoma |
| BRCA-LumA | Breast invasive carcinoma-Luminal A |
| BRCA-LumB | Breast invasive carcinoma-Luminal B |
| CESC | Cervical squamous cell carcinoma and endocervical adenocarcinoma |
| CHOL | Cholangiocarcinoma |
| COAD | Colon adenocarcinoma |
| COADREAD | Colon adenocarcinoma/Rectum adenocarcinoma Esophageal |
| DLBC | Lymphoid Neoplasm Diffuse Large B-cell Lymphoma |
| ESCA | Esophageal carcinoma |
| GBM | Glioblastoma multiforme |
| GBMLGG | Glioblastoma and Lower Grade Glioma |
| HNSC | Head and Neck squamous cell carcinoma |
| HNSC+HPV | HPV positive HNSC |
| KICH | Kidney Chromophobe |
| KIPAN | Pan-kidney cohort (KICH+KIRC+KIRP) |
| KIRC | Kidney renal clear cell carcinoma |
| KIRP | Kidney renal papillary cell carcinoma |
| LAML | Acute Myeloid Leukemia |
| LGG | Brain Lower Grade Glioma |
| LIHC | Liver hepatocellular carcinoma |
| LUAD | Lung adenocarcinoma |
| LUSC | Lung squamous cell carcinoma |
| MESO | Mesothelioma |
| OV | Ovarian serous cystadenocarcinoma |
| PAAD | Pancreatic adenocarcinoma |
| PCPG | Pheochromocytoma and Paraganglioma |
| PRAD | Prostate adenocarcinoma |
| READ | Rectum adenocarcinoma |
| SARC | Sarcoma |
| STAD | Stomach adenocarcinoma |
| SKCM | Skin Cutaneous Melanoma |
| SKCM-P | Skin Cutaneous Melanoma- Primary |
| SKCM-M | Skin Cutaneous Melanoma- Metastatic |
| STES | Stomach and Esophageal carcinoma |
| TGCT | Testicular Germ Cell Tumors |
| THCA | Thyroid carcinoma |
| THYM | Thymoma |
| UCEC | Uterine Corpus Endometrial Carcinoma |
| UCS | Uterine Carcinosarcoma |
| UVM | Uveal Melanoma |
| OS | Osteosarcoma |
| ALL | Acute Lymphoblastic Leukemia |
| ALL-R | Acute Lymphoblastic Leukemia with Resistance |

| | NB | Neuroblastoma |
| --- | --- | --- |
| | WT | High-Risk Wilms Tumor |

Table 2. Total 8 putative sites were predicted with relative profile score threshold 90% in JASPAR.

| Name | Score | Relative Score | Sequence ID | Start | End | Strand | Predicted Sequence |
| --- | --- | --- | --- | --- | --- | --- | --- |
| MA0806.1.TBX4 | 13.652736 | 1.000000002 | NC_000005.10:132816876-132818786 | 1688 | 1695 | + | AGGTGTGA |
| MA0806.1.TBX4 | 12.503169 | 0.976821265 | NC_000005.10:132816876-132818786 | 530 | 537 | + | AGGTGTTA |
| MA0593.2.FOXP2 | 11.231367 | 0.937380309 | NC_000005.10:132816876-132818786 | 244 | 252 | + | AGAAAACAG |
| MA0806.1.TBX4 | 10.179199 | 0.929963011 | NC_000005.10:132816876-132818786 | 744 | 751 | - | AGGCGTGA |
| MA0593.2.FOXP2 | 10.33313 | 0.921105532 | NC_000005.10:132816876-132818786 | 537 | 545 | + | AGAAAACAC |
| MA0593.1.FOXP2 | 11.4998 | 0.918629804 | NC_000005.10:132816876-132818786 | 243 | 253 | + | TAGAAAACAGA |
| MA0806.1.TBX4 | 8.947575 | 0.905129738 | NC_000005.10:132816876-132818786 | 642 | 649 | - | AGGTGATA |
| MA0593.2.FOXP2 | 9.4173355 | 0.904512644 | NC_000005.10:132816876-132818786 | 391 | 399 | - | AATAAACAT |

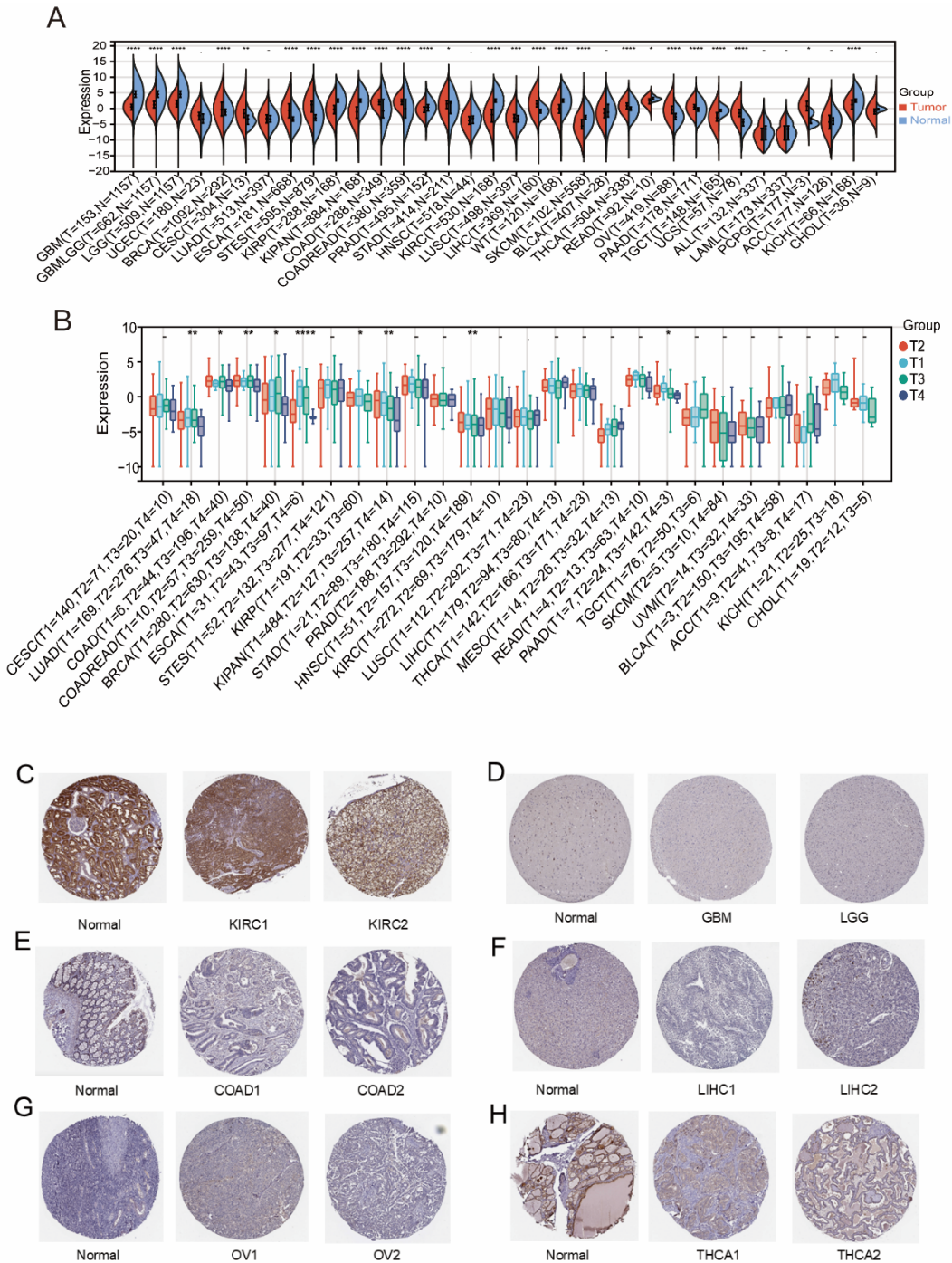

**Figure 1. (A)** *SOWAHA* exhibits highly expression levels in 14 types of tumors and is low expressed in 12 types of tumors. **(B)** The expression in different tumor stage of *SOWAHA* in 26 types of cancers and normal tissues from the TCGA and GTEx database. The expression of *SOWAHA* protein is lower in tumor than in normal tissue, such as KIRC(C), GBM, LGG (D), COAD (E), LIHC (F), OV (G), and THCA (H). The immunohistochemistry (IHC) results are obtained from the HPA database. Anti-SOWAHA antibody is HPA064621 produced in rabbit. (*p< 0.05, **p< 0.01, ***p< 0.001, ****p<0.0001)

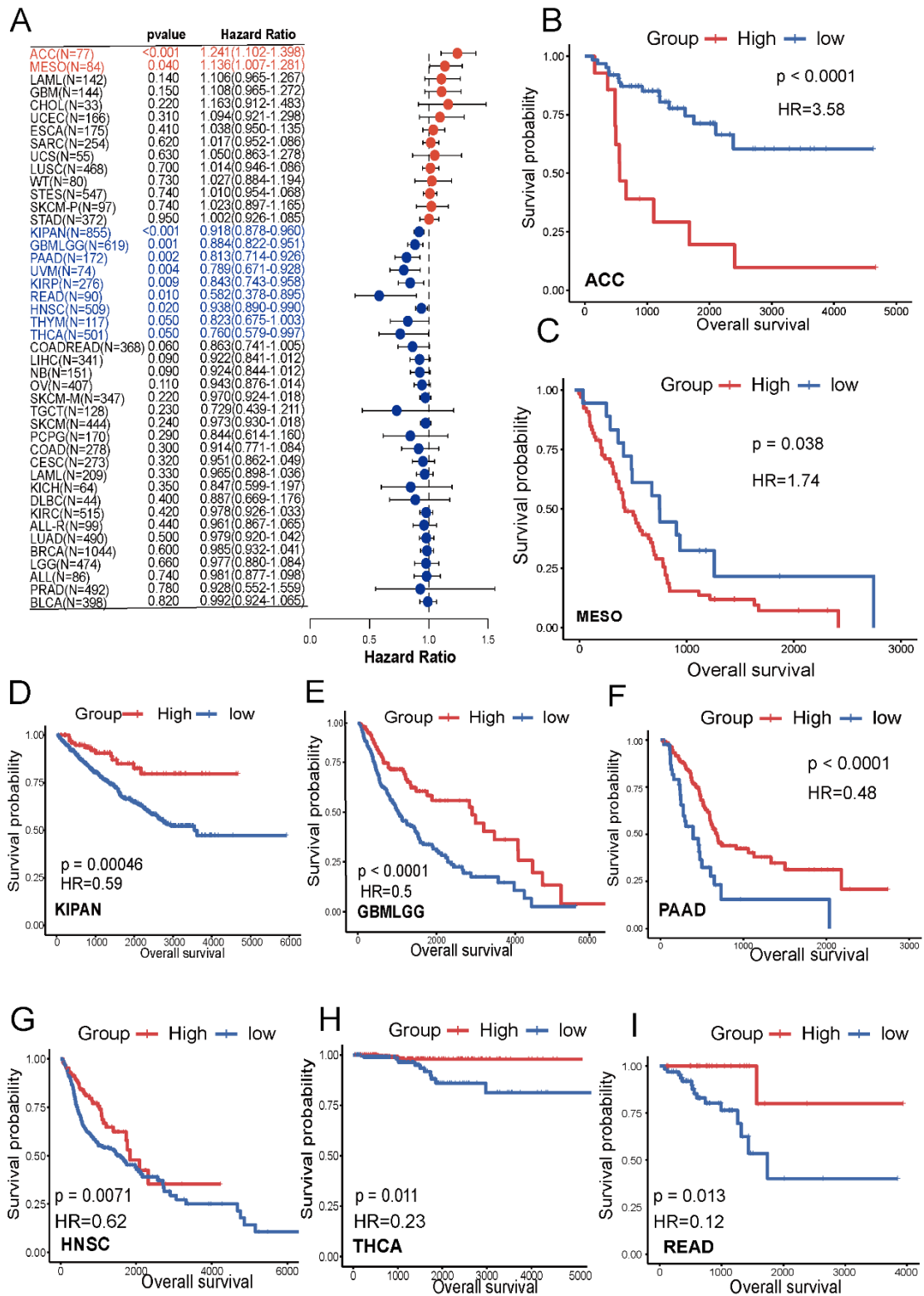

**Figure 2. (A)** *SOWAHA* expression correlates with an unfavorable prognosis of OS in 2 types of cancer patients and a favorable prognosis of OS in 9 types of cancer patients. Forest plot depicting the survival analysis results of *SOWAHA* expression on OS in pan-cancer. Results of Kaplan-Meier analysis of significance between *SOWAHA* expression and OS (p<0.05). **(B-C)** High expression of *SOWAHA* is correlated with an unfavorable prognosis of OS in ACC and MESO, **(D-I)** High

expression of *SOWAHA* is correlated with a favorable prognosis of OS in KIPAN, GBMLGG, PAAD, HNSC, THCA, and READ.

**Figure 3. Genetic alterations of *SOWAHA* in pan-cancer.** (A) Genetic alterations of *SOWAHA* protein structural domains. (B) Mutation types and frequency of *SOWAHA* gene in pan-cancer. (C) The CNA types of *SOWAHA* in 30 types of cancer. (*p < 0.05, **p < 0.01, ***p< 0.001, ****p< 0.0001)

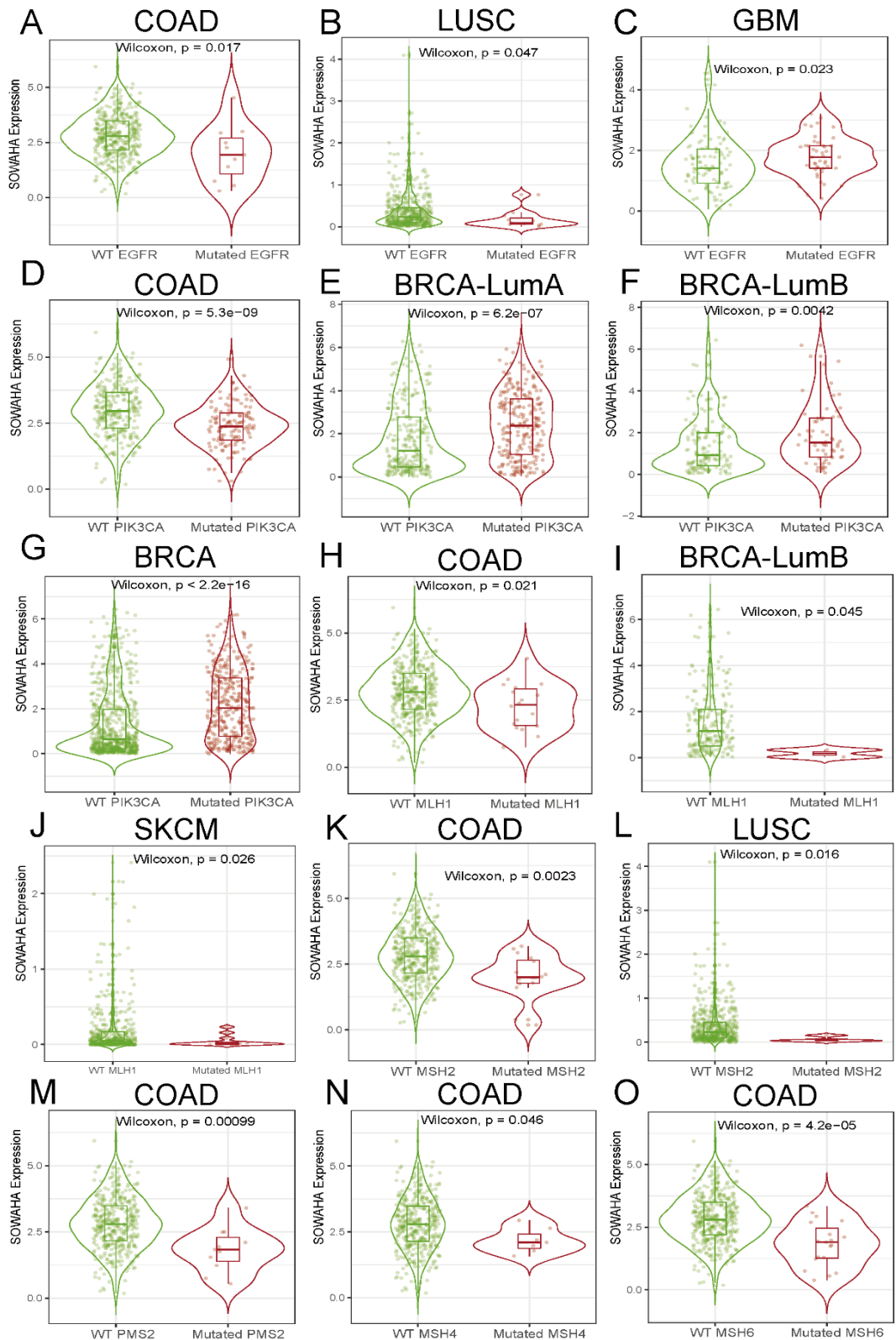

**Figure 4.** (A-B) Reduced expression of *SOWAHA* in COAD and LUSC with mutated EGFR. (C) Increased expression of *SOWAHA* in GBM with mutated EGFR. (D-F) Increased expression of SOWAHA in BRCA with mutated PIK3CA. (H-J) Reduced expression of *SOWAHA* in COAD, BRCA-LumB, and SKCM with mutated MLH1. (K-L) Reduced expression of *SOWAHA* in COAD, and LUSC with mutated MSH2. (M) Reduced expression of *SOWAHA* in COAD with mutated PMS2. (N) Reduced expression of *SOWAHA* in COAD with mutated MSH4. (O) Decreased expression of SOWAHA in COAD with mutated MSH6.

**Figure 5.** (A) Promoter methylation levels of *SOWAHA* promotor in pan-cancer. (*p < 0.05, **p < 0.01, ***p< 0.001, ****p< 0.0001) Results of Kaplan-Meier analysis of significance between *SOWAHA* expression and DNA methylation. (B-D) Significant negative correlation between 1stExon-island-cg02699218 methylation of *SOWAHA* and survival probability of ACC, MESO,

and READ. (E-G) Significant negative correlation between 1stExon-island-cg01435564 methylation of *SOWAHA* and survival probability of ACC, MESO, and READ. (H-J) Significant negative correlation between 1stExon-island-cg04044664 of *SOWAHA* and survival probability of ACC, MESO, and READ.

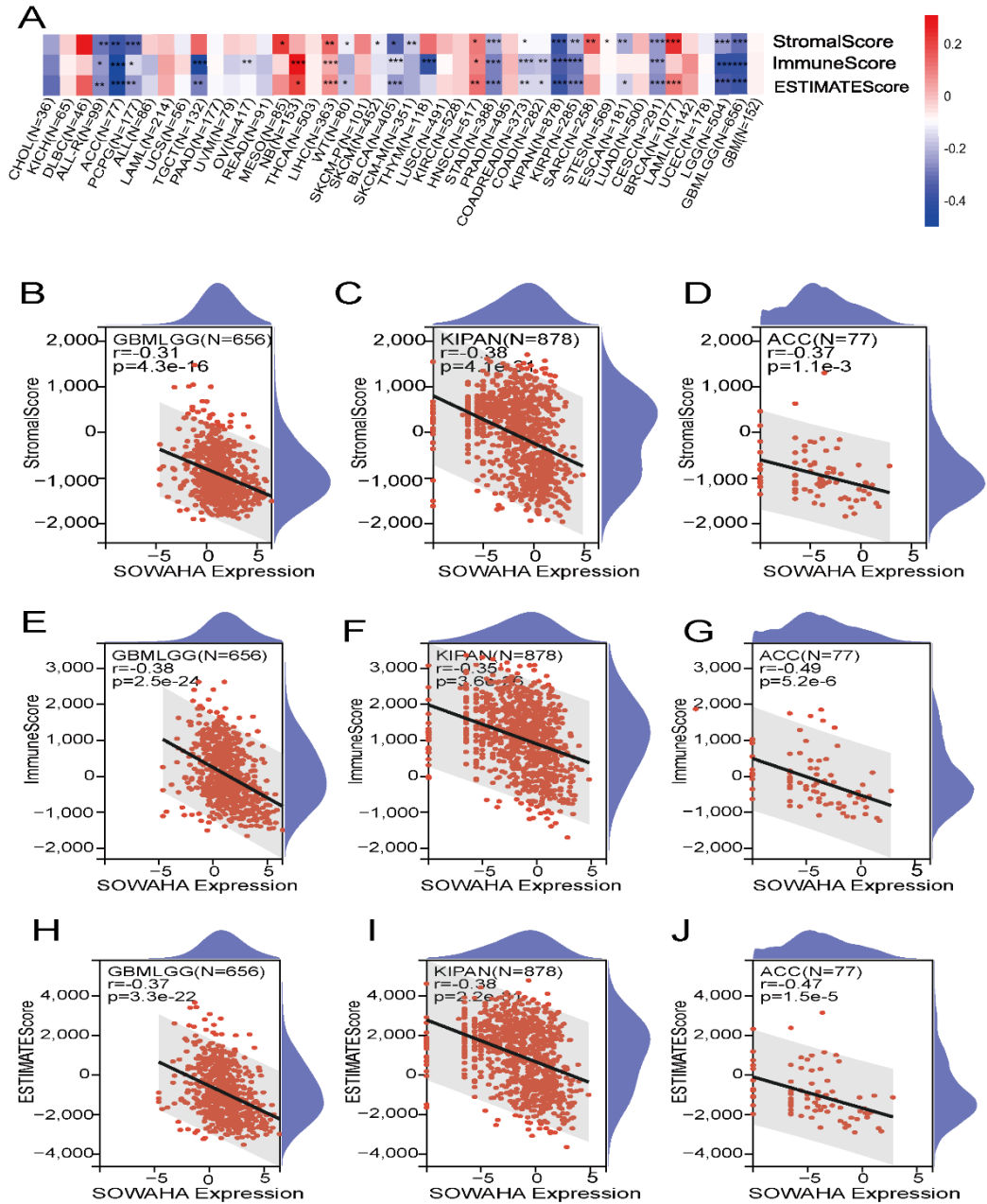

**Figure 6. *SOWAHA* expression is associated with stromal scores, immune scores, and estimate scores in the TIME across various cancer types.** (A) Heat map showing the correlation between *SOWAHA* in pan-cancer and StromalScore, ImmuneScore, and ESTIMATEScore, red represents positive correlation, and blue represents negative correlation. (B-D) Specific correlation plots between *SOWAHA* expression in GBMLGG, KIPAN, ACC and StromalScore. (E-G) Specific correlation plots between *SOWAHA* expression in GBMLGG, KIPAN, ACC and ImmuneScore. (H-J) Specific

correlation plots between *SOWAHA* expression in GBMLGG, KIPAN, ACC and ESTIMATEScore. (*p <0.05, **p< 0.01, ***p < 0.001)

**Figure 7**. (A) Association between *SOWAHA* expression and TIIC across 30 types of cancer. (B) Association between *SOWAHA* copy number alternation (CNA) and TIIC. (C) Association between *SOWAHA* Methylation and TIIC. (D) Association between *SOWAHA* expression and

MHC across 30 types of cancer. (E) Association between CNA of *SOWAHA* and MHC. (F) Association between *SOWAHA* Methylation and MHC. The red color indicates significant positive correlation and the blue color indicates significant negative correlation.

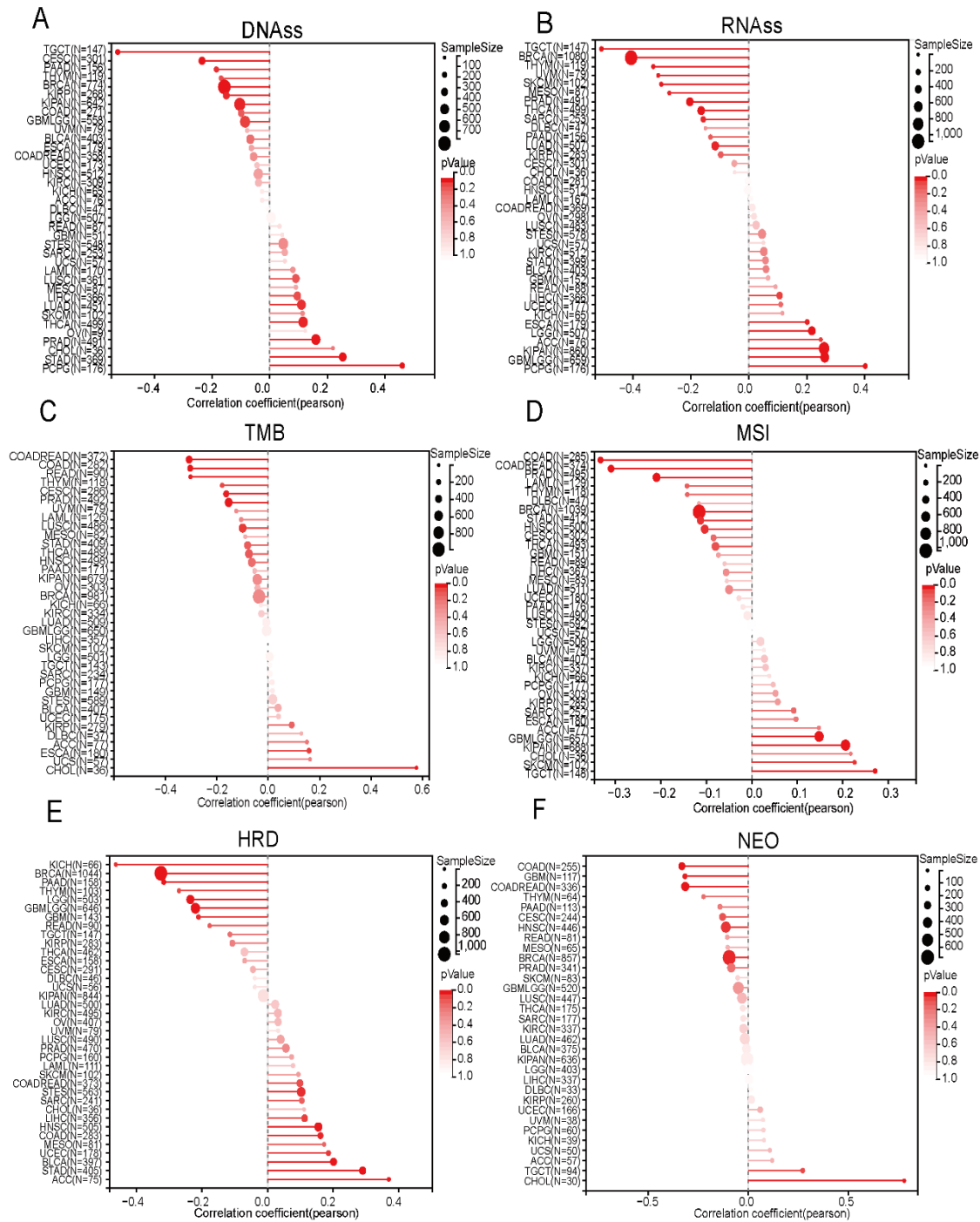

**Figure 8. The expression of *SOWAHA* is significantly associated with Tumor heterogeneity and tumor stemness in multiple types of cancer.** (A)Lollipop Chart of the correlation between *SOWAHA* expression and DNAss. (B) Lollipop Chart of the correlation between *SOWAHA* expression and RNAss. (C) Lollipop Chart of the correlation between *SOWAHA* expression and TMB. (D) Lollipop Chart of the correlation between *SOWAHA* expression and MSI. (E) Lollipop

Chart of the correlation between *SOWAHA* expression and HRD. (F) Lollipop Chart of the correlation between *SOWAHA* expression and NEO.

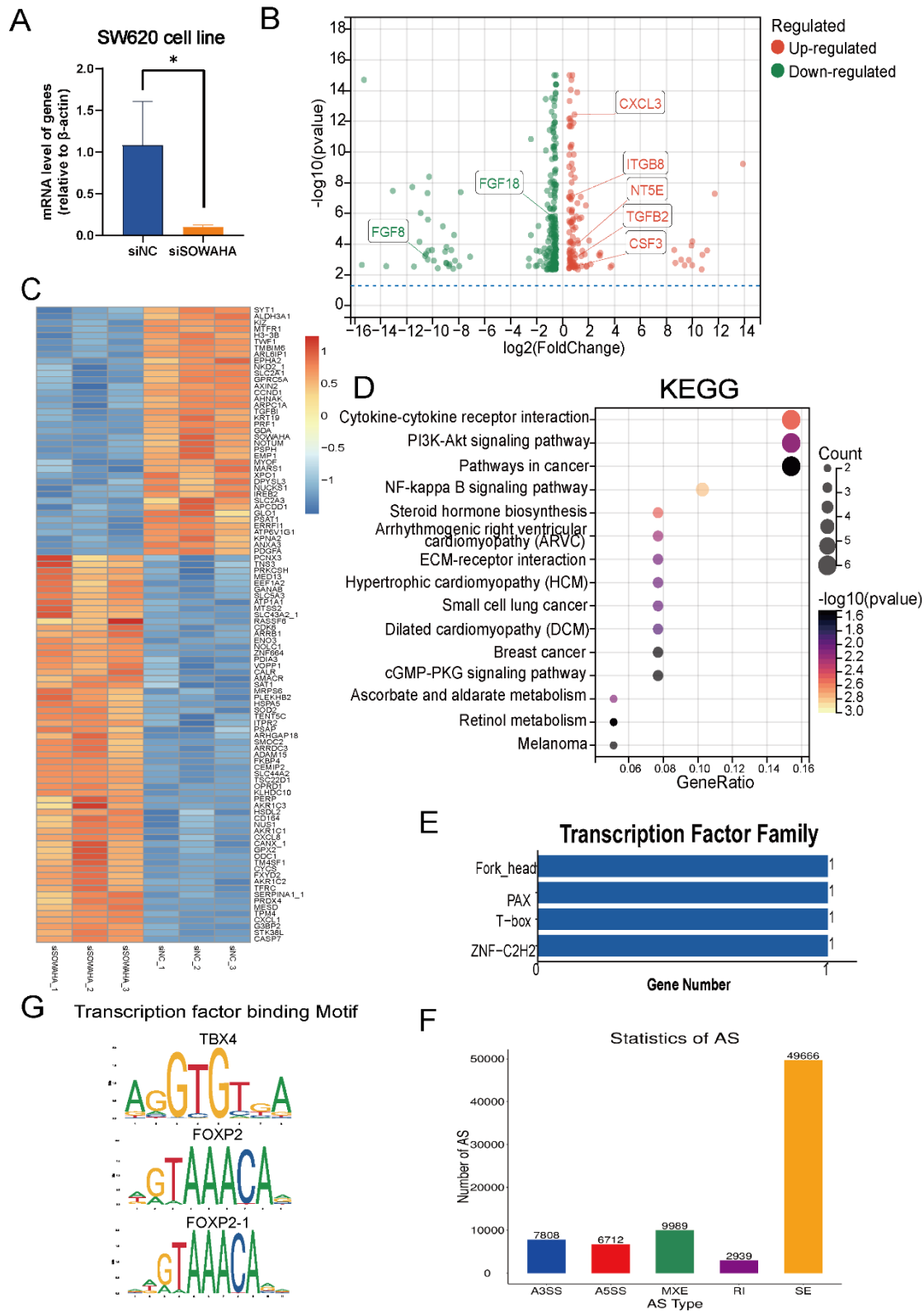

**Figure 9.  RNA-seq analysis of *SOWAHA* Knockdown in SW620 cells.** (A) The efficiency of *SOWAHA* knockdown in SW620 cells detected by q-PCR. (B) The heatmap of top 100

significantly DEGs s in SW620 cells compared to control samples (p−adj < 0.01). The red color indicates upregulation, and the green color indicates downregulation. Each row represents one gene. (C) The volcano plot shows the log2 fold-change (x-axis) versus the significance (y-axis) of all DEGs, red represents high expression and green represents low expression. (D) All significantly terms of KEGG pathway analysis of DEGs. (E) The 4 transcription factors family with significant differential expression. (F) The AS types and sites number. (G) Predictional motif of transcriptional factor bingding to *SOWAHA* promoter.

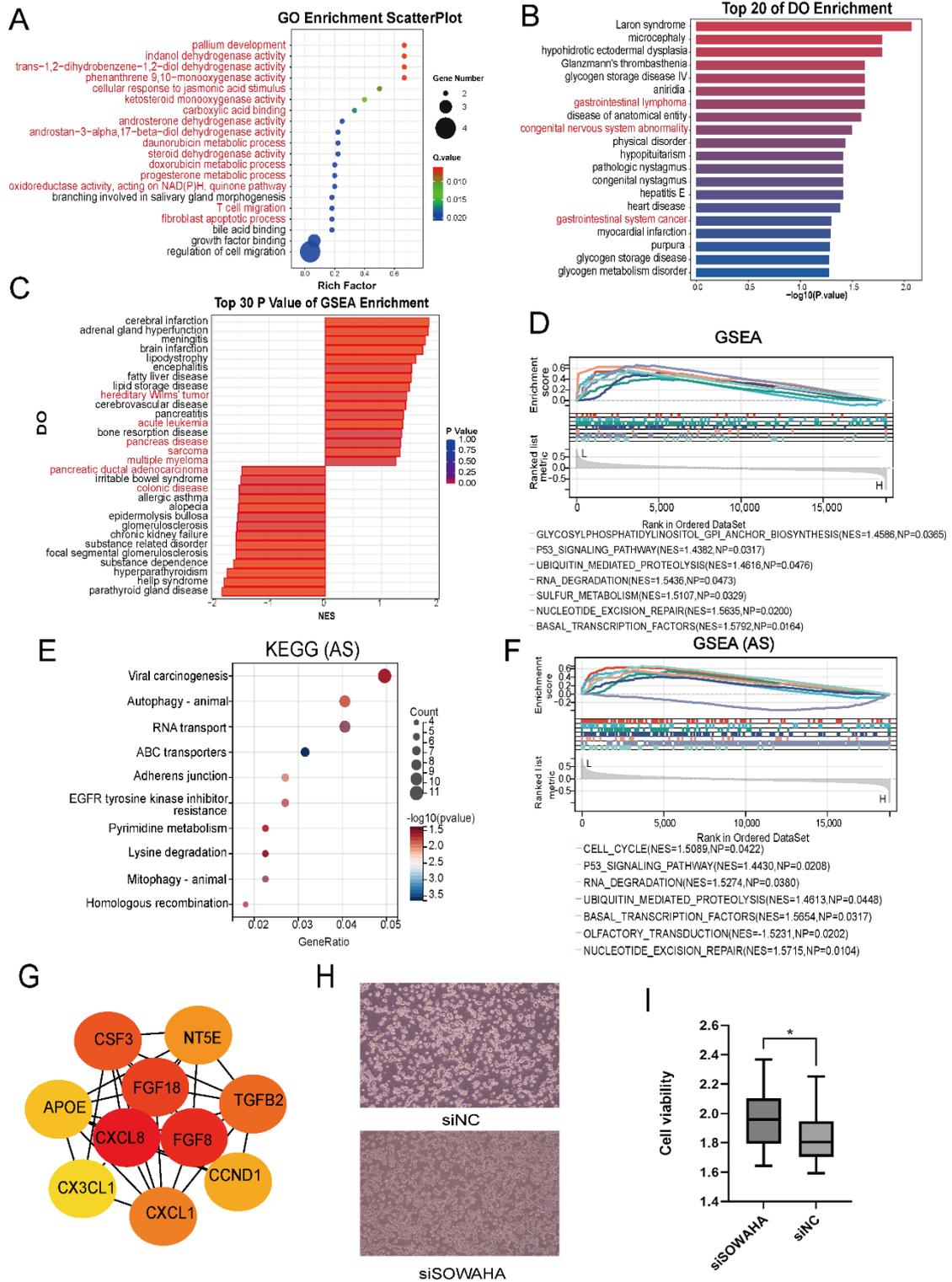

**Figure 10.** (A) The top 20 terms of GO analysis results. (B) The bar graph of DO analysis results. (C) GSEA enrichment analysis results based DO. (D) GSEA enrichment analysis results based KEGG showed all enriched significant pathways.

(E) KEGG enrichment analysis of the differentially AS genes revealed that all enriched significant pathways. (F) GSEA enrichment analysis of the differentially AS genes. (G) The 10 hub genes of PPI network. (H) The morphological changes of si*SOWAHA* and siNC samples. (I) CCK-8 assay results showed cell viability of si*SOWAHA* groups and siNC groups.

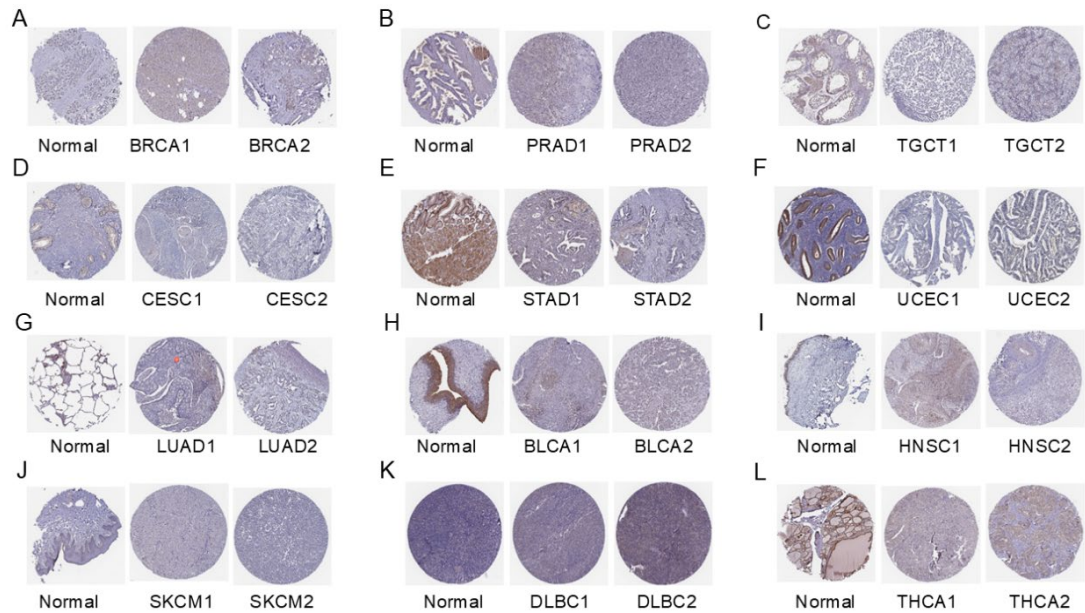

**Figure S1.** (A) The RNA expression of *SOWAHA* in cancer cell lines. The expression of SOWAHA protein is lower in tumors(right) than in normal tissue (left) in BRCA (B), PRAD(C), TGCT (D), CESC(E), STAD (F), UCEC (G), LUAD (H), BLCA(I), HNSC(J), SKCM(K), DLBC(L), and THCA(M).

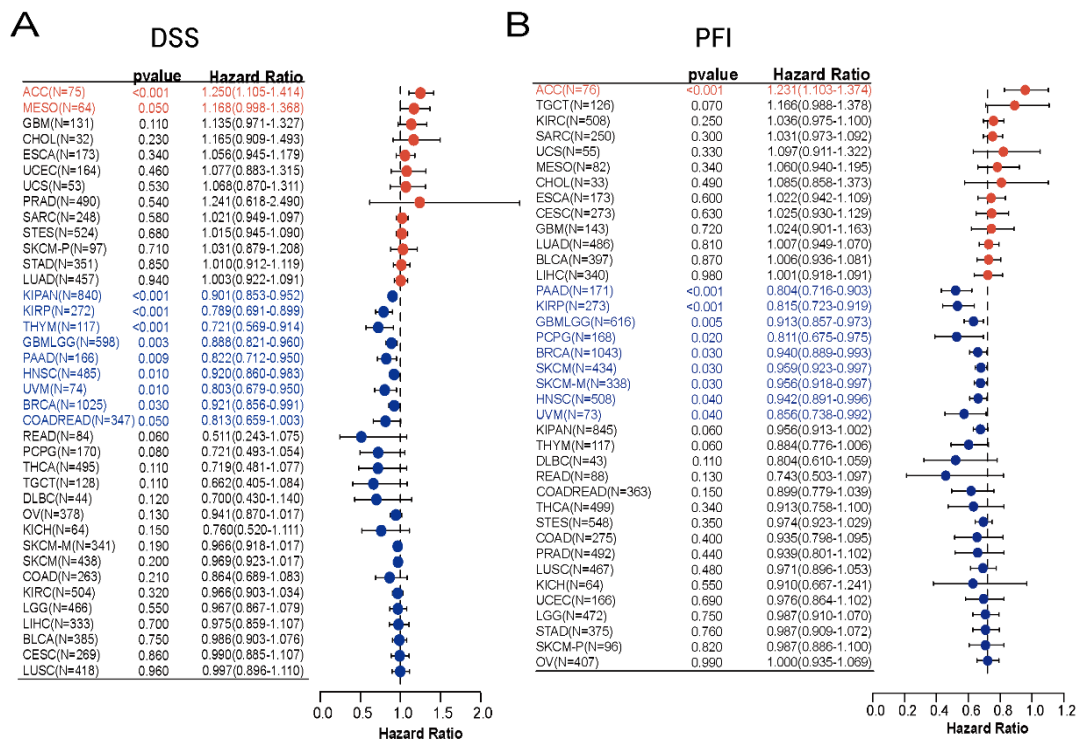

**Figure S2. Forest plot depicting the survival analysis results of *SOWAHA* expression on DSS and PFI in pan-cancer.** (A) Low expression of *SOWAHA* is significantly correlated with an unfavorable prognosis of DSS in 9 types of cancer patients. (B) Low expression of *SOWAHA* is significantly correlated with an unfavorable prognosis of PFI in 9 types of cancer patients. (p<0.05)

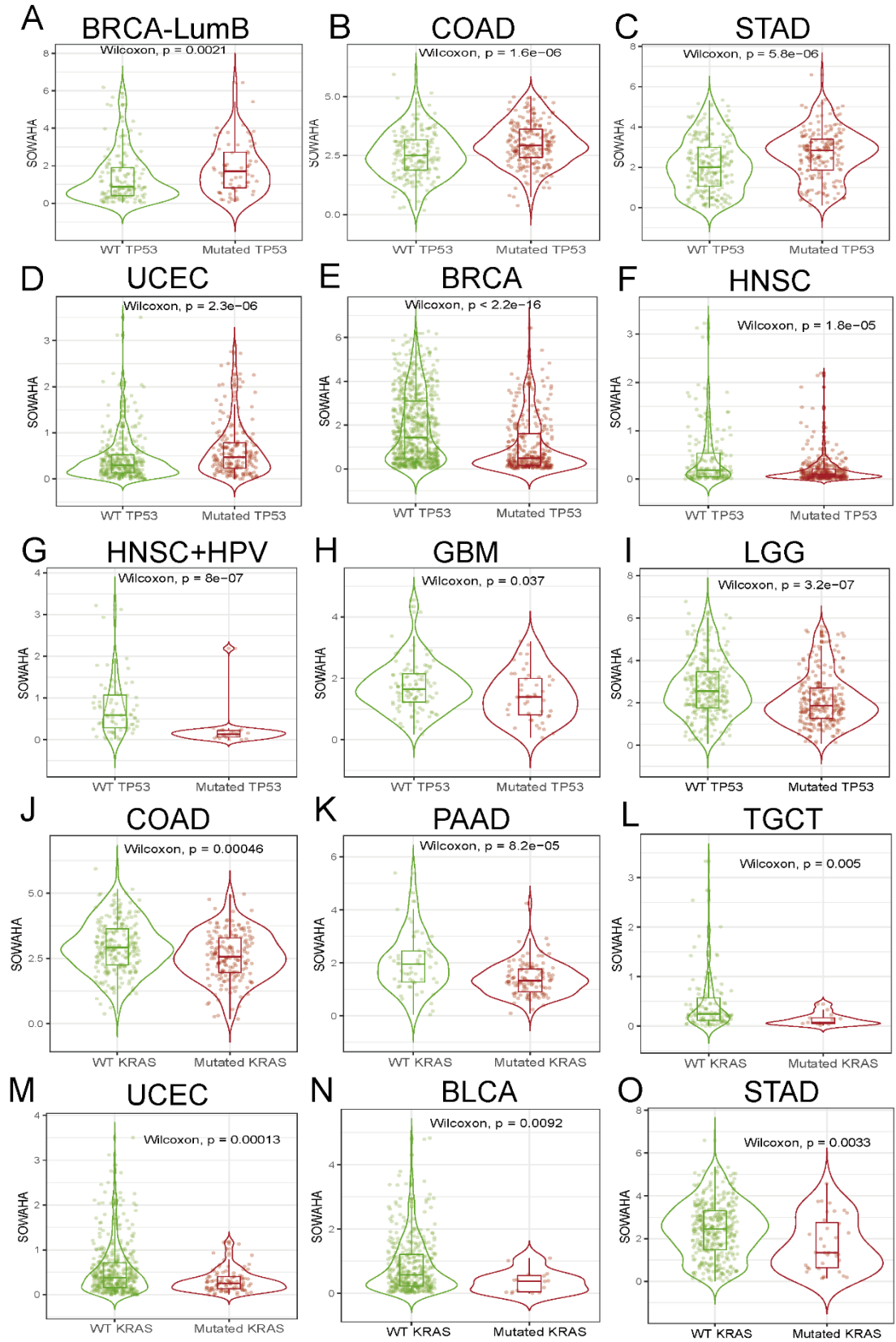

**Figure S3.** (A-D) Increased expression of *SOWAHA* in BRCA-LumB, COAD, STAD, and UCSC with mutated TP53. (E-I) Reduced expression of *SOWAHA* in mutated EGFR type of BRCA,

HNSC, HNSC+HPV, and GBM, LGG. (D-F) Decreased expression of *SOWAHA* in KRAS type of COAD, PAAD, TGCT, UCEC, BLCA, and STAD.

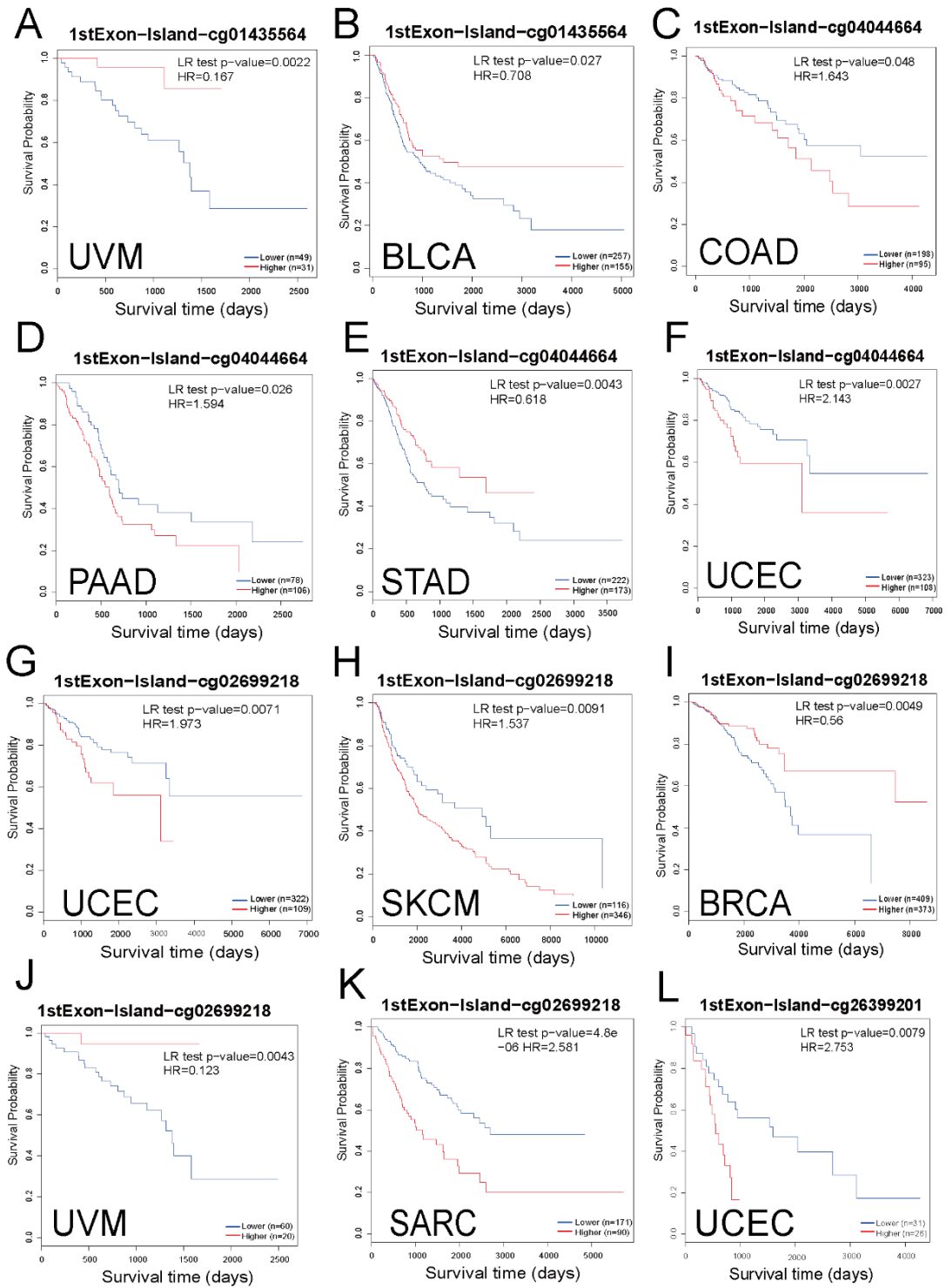

**Figure S4. The relationship between methylation of the 1stExon-Islands in *SOWAHA* and survival probability in pan-cancer.** The DNA methylation of 1stExon-Island cg01435564(A-B), 1stExon-Island cg04044664(C-F), 1stExon-Island cg02699218(G-K), 1stExon-Island cg26399201 (L).

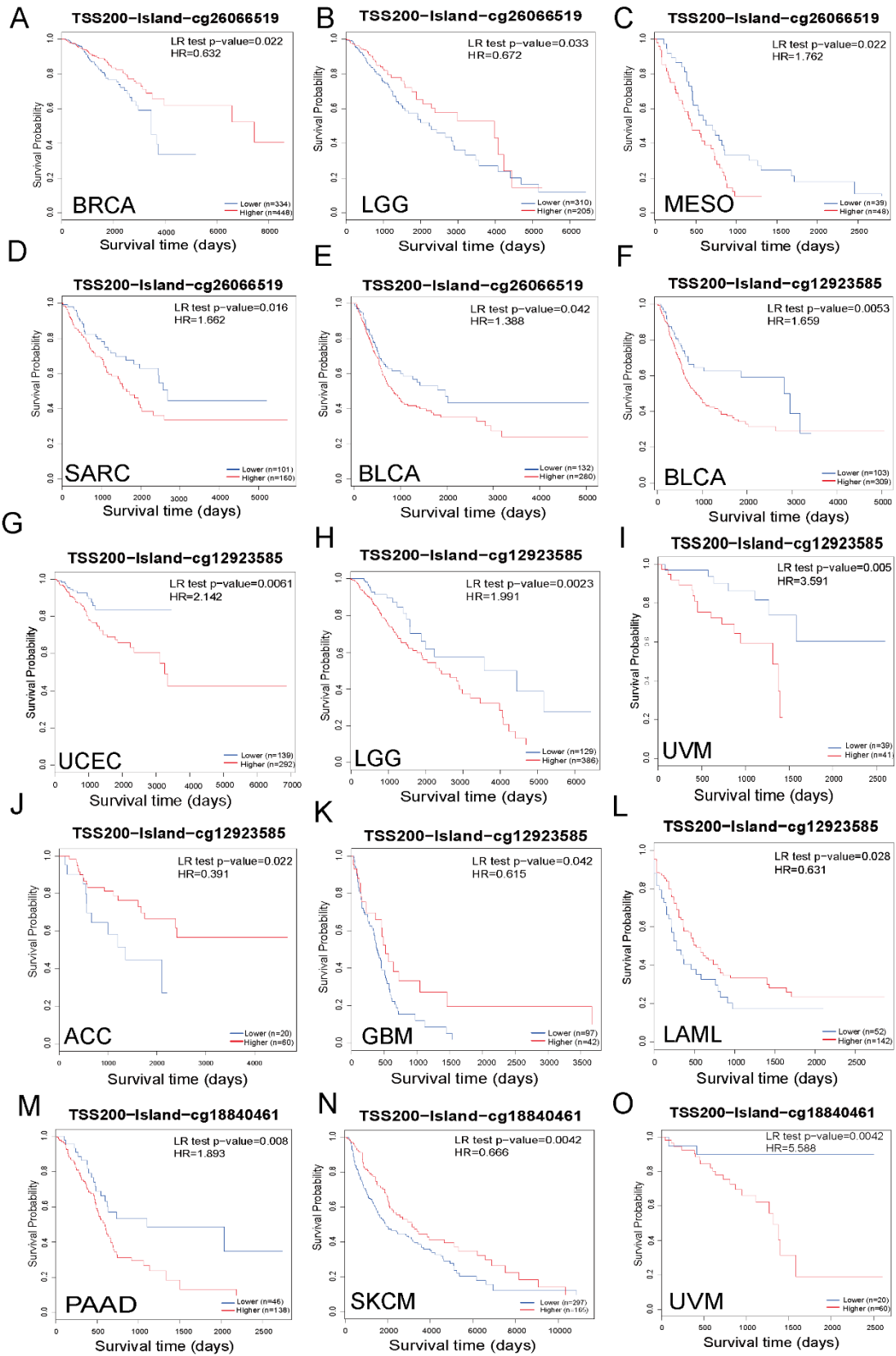

**Figure S5. The relationship between methylation of the TSS200-Islands in *SOWAHA* and survival probability in pan-cancer.** The DNA methylation of TSS-200 island cg26066519(A-E), cg12923585(F-L), cg18840461(M-O).

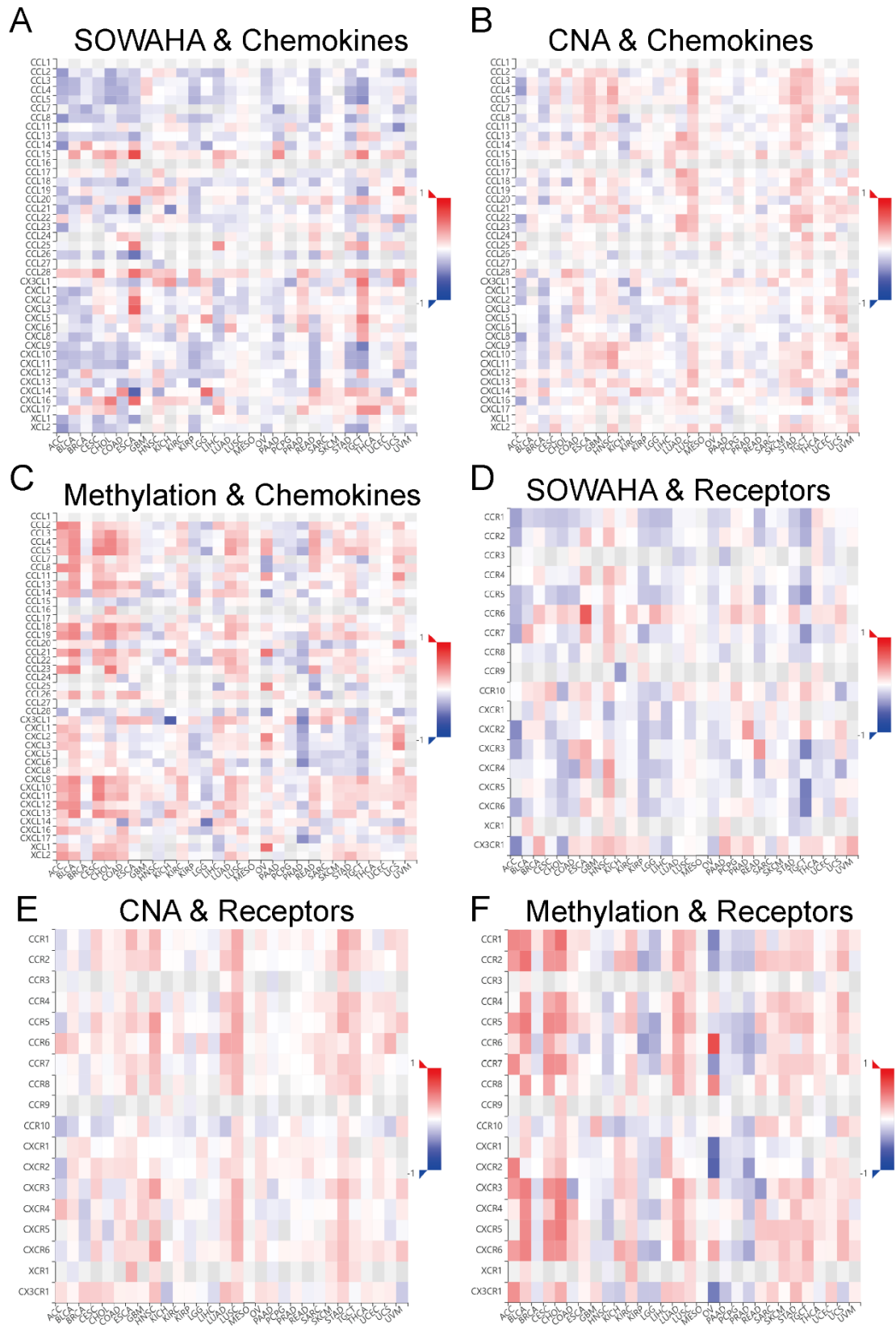

**Figure S6.** (A-C) Correlation between chemokines and *SOWAHA* Expression, CNA of *SOWAHA*, and Methylation of *SOWAHA* in pan-cancer. (D-F) Correlation between chemokine receptors and *SOWAHA* Expression, CNA of *SOWAHA*, and Methylation of *SOWAHA* in pan-cancer.

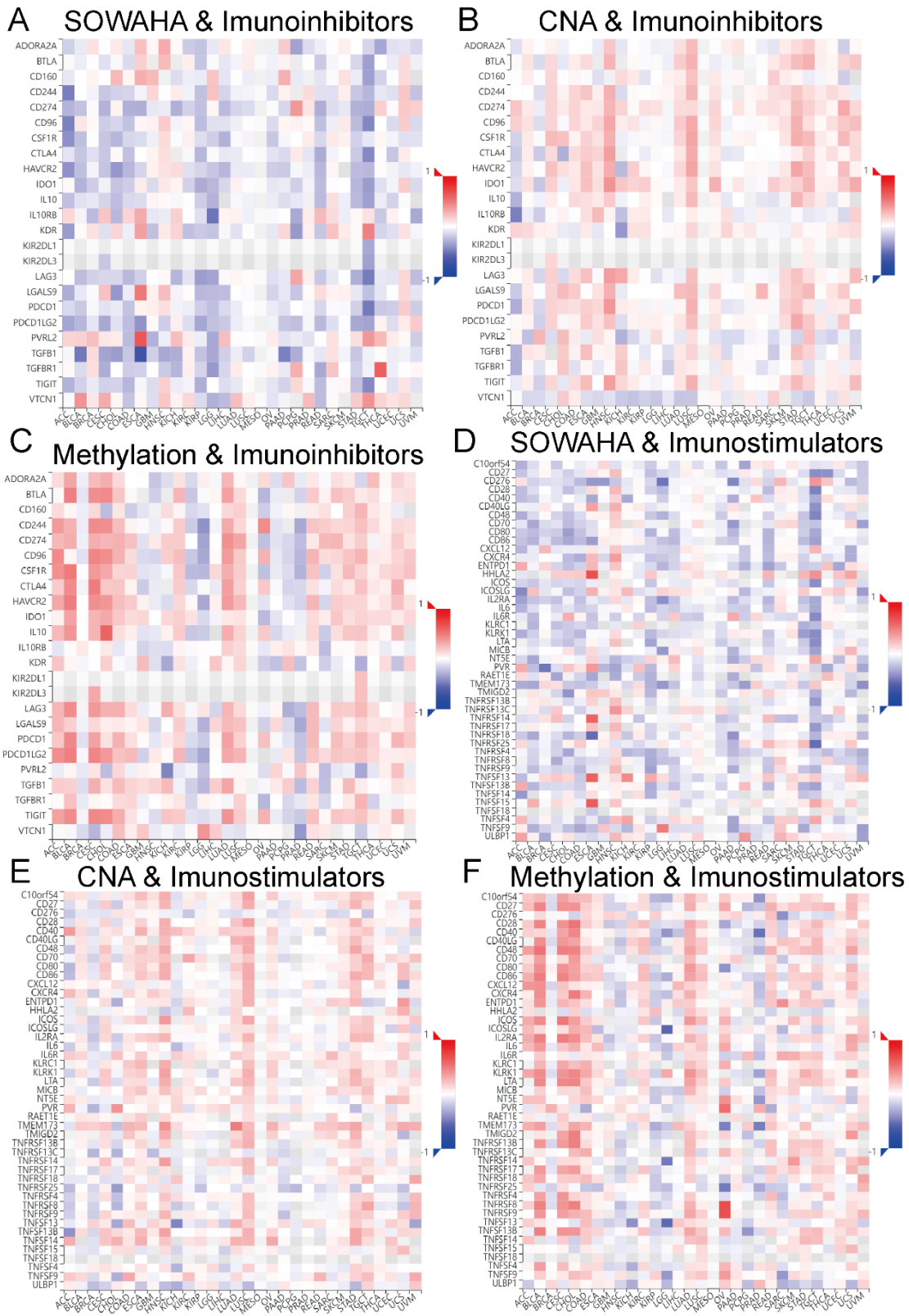

**Figure S7.** Correlation between *SOWAHA* and 69 immunomodulators (inhibitors and stimulators) across 30 types of cancer. (A) Association between *SOWAHA* and 24 immunoinhibitors across 30 types of cancer. (B) Association between CNA of *SOWAHA* and 24 immunoinhibitors across 30 types of cancer. (C) Association between Methylation of *SOWAHA* and 24 immunoinhibitors across 30 types of cancer. (D) Association between *SOWAHA* and 24 immunostimulators across 30

types of cancer. (E) Association between CNA of *SOWAHA* and 24 immunostimulators across 30 types of cancer. (F) Association between Methylation of *SOWAHA* and 24 immunostimulators across 30 types of cancer.